\documentclass[letterpaper,twocolumn,10pt]{article}

\PassOptionsToPackage{table}{xcolor}
\usepackage{usenixsecurity2026_latex_templates/usenix}

\usepackage{amsmath}
\usepackage{amssymb}
\usepackage{booktabs}

\usepackage{graphicx}
\usepackage{float}
\usepackage{caption}
\usepackage{placeins}
\usepackage{array}
\usepackage{subcaption}
\usepackage[linesnumbered,ruled,vlined]{algorithm2e}
\definecolor{lightgray}{gray}{0.45}
\SetCommentSty{\color{lightgray}\ttfamily}
\SetNlSty{texttt}{\color{black}}{}
\usepackage{multirow}
\usepackage{pifont}
\usepackage{soul}

\begin{document}

\date{}

\title{Purify Once, Edit Freely: Breaking Image Protections under Model Mismatch}

\author{
\rm Qichen Zhao$^{1}$, \quad Shengfang Zhai$^{2,\dag}$, \quad Xinjian Bai$^{1}$, \quad Qingni Shen$^{1,\dag}$, \quad Qiqi Lin$^{1}$,\\
\rm Yansong Gao$^{3}$, \quad Zhonghai Wu$^{1}$\\
{$^1$Peking University, $^2$National University of Singapore, }\\
{ \quad $^3$The University of Western Australia}\\
{\tt \{z24, baixinjian, qiqilin\}@stu.pku.edu.cn, }\\
{\tt \{qingnishen, wuzh\}@pku.edu.cn, shengfang.zhai@nus.edu.sg, }\\
{\tt gao.yansong@hotmail.com}
}

\maketitle

{
\renewcommand{\thefootnote}{\fnsymbol{footnote}}
\footnotetext[2]{Corresponding authors.}
}

\begin{abstract}
Diffusion models enable high-fidelity image editing, but they can also be misused for unauthorized style imitation and harmful content generation.
To mitigate such risks, proactive image protection methods embed small, often imperceptible adversarial perturbations into images before sharing, aiming to disrupt downstream editing and fine-tuning.
In realistic post-release settings, content owners cannot control downstream reprocessing or the choice of diffusion-based editing pipelines.
Consequently, protections optimized against a surrogate model may break down when attackers purify or edit images using a mismatched pipeline built on a different model.
Prior purification methods can weaken protections, but they often trade off effective perturbation removal against preserving image utility, and they rarely evaluate model architectural mismatch explicitly.

We introduce a unified purification framework after release to evaluate protection survivability under model mismatch.
We propose two practical purifiers: \textbf{VAE-Trans}, which corrects protected images via latent-space projection, and \textbf{EditorClean}, which performs instruction-guided reconstruction using a Diffusion Transformer to leverage architectural heterogeneity.
Both operate without access to protected images or defense internals.
We conduct extensive experiments on 2,100 editing tasks across six representative protection methods.
\textbf{EditorClean} consistently restores editability.
Compared to unpurified protected inputs, it improves PSNR by 3--6~dB and reduces FID by 50--70\% on downstream edits. Compared to prior purification baselines, it typically achieves an additional about 2dB PSNR gain and 30\% lower FID.
Our findings reveal a largely \emph{purify-once, edit-freely} failure mode: once an attacker succeeds in purification, the protective signal is largely erased, enabling subsequent unrestricted editing.
This is primarily due to the limited cross-model transferability of perturbations and the reconstructive nature of diffusion-based pipelines operating with heterogeneous models.
Overall, these results highlight the importance of evaluating robustness under model mismatch and the need to design adversarial protections that remain effective against heterogeneous attackers.
\end{abstract}

\section{Introduction}

Diffusion models have become a foundational technique for image synthesis and editing~\cite{sohl2015deep,ho2020denoising,rombach2022high}, enabling high-fidelity semantic manipulation, inpainting, and style transfer from natural language instructions~\cite{meng2021sdedit,hertz2022prompt,brooks2023instructpix2pix,zhang2023adding}. These capabilities substantially expand creative and editing workflows, but they also introduce new avenues for misuse, such as unauthorized style imitation and harmful content generation. As a result, protecting visual content against downstream misuse has become an increasingly important security concern.
Proactive image protection has emerged as a promising defensive strategy to address this challenge~\cite{salman2023raising,shan2023glaze,liang2023mist,liang2023adversarial,van2023anti,choi2024diffusionguard}. By embedding small and often imperceptible adversarial perturbations into images prior to release, these methods aim to disrupt subsequent generative editing or model fine-tuning. However, once images are released, they may be further processed or modified by users in uncontrolled ways~\cite{wallace1991jpeg,sandoval2023jpeg,honig2024adversarial}.

In practice, attackers can freely collect protected images and reprocess them using a broad spectrum of tools, ranging from simple image transformations to generative editing pipelines built on heterogeneous model architectures~\cite{sandoval2023jpeg,honig2024adversarial,cao2023impress,zhao2024can}. Existing protection methods are typically optimized against a specific surrogate model and evaluated in matched settings~\cite{salman2023raising,liang2023adversarial,liang2023mist,choi2024diffusionguard}.
As a result, their effectiveness is not evaluated in practice when applied across mismatched diffusion-based editing pipelines.

A key threat in this setting arises from purification attacks, in which attackers attempt to remove or attenuate protective perturbations before performing unauthorized edits or downstream fine-tuning. Prior studies~\cite{sandoval2023jpeg,honig2024adversarial,nie2022diffusion,cao2023impress,zhao2024can} show that even simple preprocessing or reconstruction operations can weaken protections. However, existing approaches often struggle to simultaneously eliminate protective signals and preserve image utility. They also rarely evaluate purification explicitly under the mismatch that matters in practice: the defender crafts perturbations against a surrogate model, but the attacker may purify the image before editing it with a different model. Together, these limitations highlight the need for a more systematic and realistic evaluation of protection survivability under purification-based attacks.

We address these gaps by proposing a unified purification framework after release for evaluating perturbation-based image protection~\cite{foerster2025lightshed}. In our formulation, defenders cannot observe or constrain the attacker-side pipeline after release, while attackers may freely choose heterogeneous models and preprocessing pipelines~\cite{honig2024adversarial,cao2023impress,zhao2024can}. Our framework models an attacker who either directly edits the protected image $x_{\mathrm{adv}}$ using a different diffusion-based editing pipeline, or prepends a purification operator $\mathcal{P}$ before editing. This formulation enables systematic measurement of protection survivability across purification strategies and model architectures by focusing on editability after release.
Our key observation is that while protective perturbations can be highly effective on the surrogate models they are optimized against, they often generalize poorly across architectures. Leveraging this observation, we propose two practical purification methods to instantiate the framework. \textbf{VAE-Trans} fine-tunes a VAE encoder to correct protected images through latent-space projection, probing non-robustness to distribution shift within the same model family. \textbf{EditorClean} formulates purification as instruction-guided reconstruction using a Diffusion Transformer~\cite{peebles2023scalable}, adapting the ICEdit in-context editing framework~\cite{zhang2025enabling} to adversarial purification. It exploits architectural mismatch by leveraging the limited transferability of perturbations optimized for a surrogate model, showing that model mismatch can already be effective in practical scenarios.

Both methods are trained solely on public data and require no access to protected images or defense internals, making them practical in realistic settings after release. Following the DiffusionGuard evaluation protocol~\cite{choi2024diffusionguard}, we conduct extensive experiments across six representative protection methods: PhotoGuard~\cite{salman2023raising}, AdvDM~\cite{liang2023adversarial}, MIST~\cite{liang2023mist}, SDS~\cite{DBLP:conf/iclr/XueLWC24}, DiffusionGuard~\cite{choi2024diffusionguard}, and AdvPaint~\cite{jeon2025advpaintprotectingimagesinpainting}. We evaluate 2,100 image editing tasks spanning celebrity portraits, objects, and animals.
Our experiments show that purification under model mismatch can substantially restore editability. Compared with unpurified protected inputs, \textbf{EditorClean} improves PSNR by about 3--6~dB and reduces FID by about 50--70\% across all six defenses. Relative to existing purification baselines, it further boosts PSNR by about 1--3~dB and lowers FID by 10--40\%, substantially narrowing the gap to clean-image edits.
Our experiments reveal a severe post-release failure mode of current proactive image protection: once an image is successfully purified, the protective signal is largely removed, enabling unrestricted downstream editing thereafter.

In summary, this work makes the following contributions:
\begin{itemize}
\item We show that protective perturbations transfer poorly across heterogeneous models, and introduce a benchmark that comprehensively evaluates the protection effectiveness under model mismatch.

\item We propose \textbf{VAE-Trans} and \textbf{EditorClean}, two practical purifiers trained on public data. We show they substantially restore the editability of protected images and can further restore the utility of unlearnable examples.

\item We highlight the practical vulnerability of protective perturbations where heterogeneous models effectively act as purifiers, enabling attackers to eliminate protections and then edit freely, motivating more realistic evaluation and defense design under model mismatch.
\end{itemize}

\section{Related Work}

\subsection{Image Generation and Editing}

Diffusion probabilistic models~\cite{sohl2015deep,ho2020denoising,dhariwal2021diffusion,croitoru2023diffusion} have emerged as the dominant paradigm for high-fidelity image synthesis.
Latent diffusion models~\cite{rombach2022high} reduce computational cost by operating in compressed latent space, enabling large-scale text-to-image systems such as Stable Diffusion~\cite{rombach2022high}, DALL-E 2~\cite{ramesh2022hierarchical}, and Imagen~\cite{saharia2022photorealistic}.
Beyond generation, diffusion models enable advanced editing: SDEdit~\cite{meng2021sdedit} performs noise-denoise refinement, Prompt-to-Prompt~\cite{hertz2022prompt} manipulates cross attention maps, InstructPix2Pix~\cite{brooks2023instructpix2pix} enables instruction following edits, and ControlNet~\cite{zhang2023adding} provides structural conditioning.
These advances amplify both creative potential and misuse risks.

\subsection{Proactive Image Protection}

Proactive defenses embed imperceptible adversarial perturbations into images to disrupt downstream diffusion based editing~\cite{liang2023adversarial,salman2023raising,shan2023glaze}.
Related ideas also appear in broader settings such as unlearnable examples for preventing unauthorized model training and privacy preserving perturbations against face recognition~\cite{huang2021unlearnable,fu2022robust,cherepanova2021lowkey}.
PhotoGuard~\cite{salman2023raising} attacks VAE latents or the full LDM pipeline; GLAZE~\cite{shan2023glaze} and Mist~\cite{liang2023mist} target style mimicry with perceptual constraints; Anti-DreamBooth~\cite{van2023anti} disrupts subject driven personalization.
Subsequent work improves efficiency: Lo et al.~\cite{lo2024distraction} attack text salient regions via cross attention; Xue et al.~\cite{DBLP:conf/iclr/XueLWC24} leverage score distillation sampling; DiffusionGuard~\cite{choi2024diffusionguard} introduces mask augmentation for inpainting robustness.
However, protections optimized against specific surrogates often degrade under different editing pipelines~\cite{honig2024adversarial}.

\subsection{Purification of Protective Perturbations}

Recent studies expose fragility in perturbation-based protections.
Honig et al.~\cite{honig2024adversarial} show that simple preprocessing such as Gaussian noise or upscaling can weaken protections like GLAZE and Mist.
Sandoval-Segura et al.~\cite{sandoval2023jpeg} demonstrate that JPEG compression bypasses protections by attenuating high frequency perturbations.
Cao et al.~\cite{cao2023impress} propose IMPRESS, exploiting inconsistencies between protected images and their VAE reconstructions.
Zhao et al.~\cite{zhao2024can} introduce GridPure, a patch level reconstruction approach for high resolution scenarios that adapts DiffPure~\cite{nie2022diffusion} diffusion purification to generative protection bypass.
Additionally, LightShed~\cite{foerster2025lightshed} presents a learning based depoisoning attack that leverages the availability of protection schemes to learn and subtract scheme specific perturbation patterns, demonstrating that such perturbations can exhibit exploitable regularities.

\begin{figure*}[t]
\centering
\includegraphics[width=\textwidth]{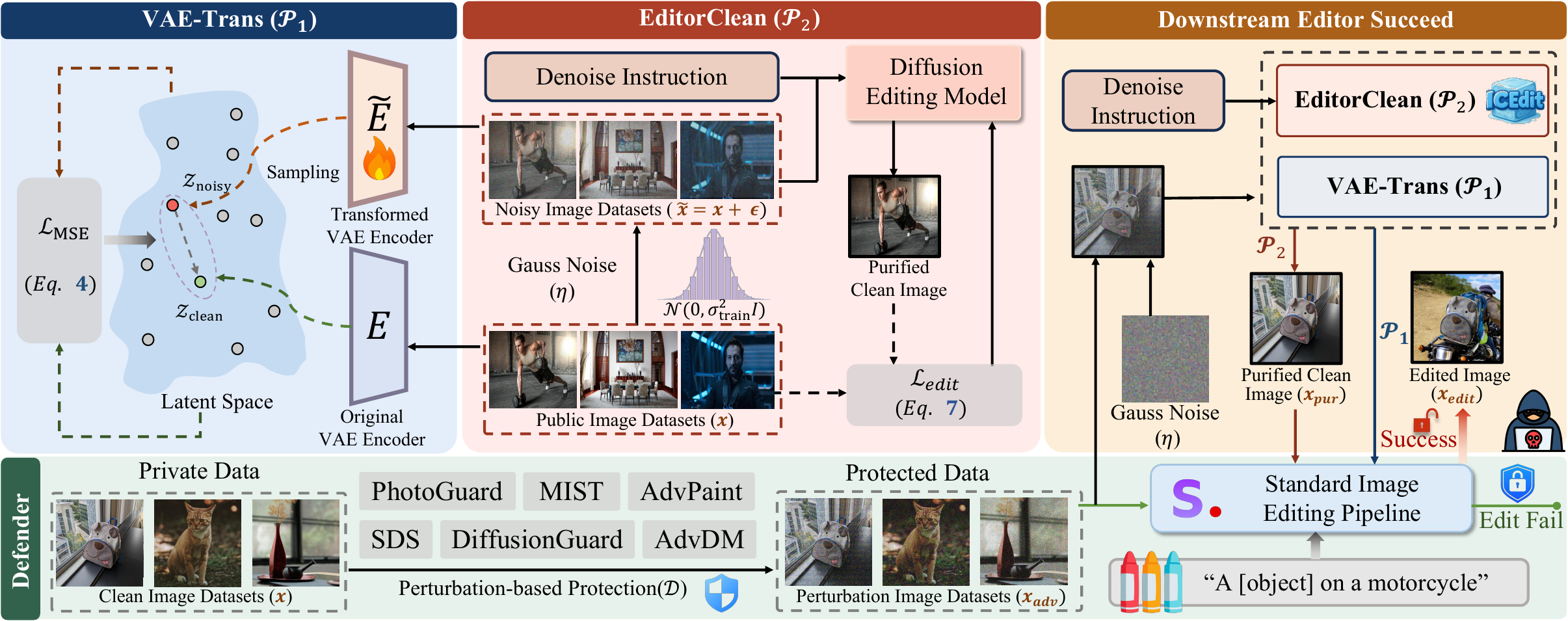}
\caption{Schematic overview of our post-release purification setting and two purifiers, \textbf{VAE-Trans} (Section~\ref{sec:vae-trans}) and \textbf{EditorClean} (Section~\ref{sec:editorclean}). The \emph{Standard image editing pipeline} in the figure denotes the defender's surrogate editor used to optimize the protective perturbation. \textbf{Left (Training):} Both purifiers are trained on public image datasets. \textbf{Right (Inference):} Given $x_{\mathrm{adv}}$, the attacker prepends a  purifier to obtain a purified image $x_{\mathrm{pur}}$; for \textbf{EditorClean}, we additionally inject light Gaussian noise before reconstruction. The purified image is then passed to a downstream editor for editing.}
\label{fig:overview}
\end{figure*}

\section{Threat Model}\label{sec:threat}

We study the security of perturbation-based image protection under a realistic deployment setting where protected images are disseminated through uncontrolled channels and may be further processed before editing~\cite{salman2023raising,liang2023mist,choi2024diffusionguard,foerster2025lightshed}.
We consider a \emph{post-release adversarial setting}. Once $x_{\mathrm{adv}}$ is published, images may undergo uncontrolled downstream processing: the defender cannot observe, restrict, or anticipate the attacker-side pipeline, including any purification steps and the choice of downstream model.
Accordingly, we distinguish (i)~defender-side limitations after release (lack of visibility and control) from (ii)~attacker-side capabilities (choice of purification $\mathcal{P}$ and editor $\mathcal{E}$).
Throughout, we use the term \emph{editor} to refer to a diffusion-based image editing pipeline (e.g., Stable Diffusion v1.5 inpainting).
Crucially, the defender crafts perturbations against a surrogate model, while the attacker may purify and edit using a different editor.
Accordingly, the attacker may either (i)~directly edit $x_{\mathrm{adv}}$ with a different editor, or (ii)~prepend a purification operator $\mathcal{P}$ and then edit the purified image.
We formalize the interaction between the defender and the attacker based on this mismatch in knowledge and capabilities, and summarize the deployment scenario, capabilities, and success criteria.

\noindent\textbf{Deployment Scenario.}
The scenario considers a standard content sharing ecosystem.
A content owner, acting as the defender, preemptively applies a protection mechanism $\mathcal{D}$ to a clean image $x$ before public release.
The resulting protected image is $x_{\mathrm{adv}} = x + \delta$.
This image is distributed through open and uncontrolled channels such as social media platforms or stock photo repositories.
Once the image is published, the defender loses all control over subsequent processing~\cite{foerster2025lightshed}.

A third party attacker can obtain $x_{\mathrm{adv}}$ (but not the original clean image $x$ or the perturbation $\delta$).
The attacker can further apply arbitrary preprocessing or purification before performing unauthorized semantic edits $y$ through a diffusion-based editing pipeline~\cite{sandoval2023jpeg,honig2024adversarial,cao2023impress,zhao2024can}.

\noindent\textbf{Attacker and Defender Capabilities.}
This deployment setting creates a fundamental imbalance in operational constraints.
The defender is proactive but constrained.
They must commit to a fixed protection strategy before release.

They must also satisfy a strict imperceptibility budget, $\|\delta\|_\infty \leq \epsilon$, to preserve visual utility.
In practice, defenders optimize perturbations against a specific surrogate model (e.g., SD~v1.5 Inpainting).
Once images are published, the attacker can switch to a different editor (e.g., a different model), making \emph{model mismatch} a core assumption in our setting.
The attacker can prepend an arbitrary purification operator $\mathcal{P}$ to the workflow~\cite{sandoval2023jpeg,honig2024adversarial,nie2022diffusion,cao2023impress,zhao2024can}.
Unlike the defender's surrogate, the attacker can exploit heterogeneous generative priors.
They can also train purification models on public and unprotected data without requiring knowledge of the defense mechanism.

\noindent\textbf{Attacker and Defender Objectives.}
The attacker follows a \emph{purify-once, edit-freely} objective.
Importantly, the ``attacker'' in our setting may be malicious or non-malicious: an end user or downstream service may simply apply off-the-shelf editing tools, unintentionally using an editor that differs from the defender's surrogate.
In this case, model mismatch and even purification-like effects can arise naturally from standard preprocessing or reconstruction steps in the downstream pipeline, without the user explicitly targeting the protection.
An attack is successful if the purified image $x_{\mathrm{pur}} = \mathcal{P}(x_{\mathrm{adv}})$ enables the target editor to produce an edited output $x_{\mathrm{edit}}$.
This output must follow the semantic instruction $y$ and achieve visual quality comparable to edits applied to the original clean image~\cite{choi2024diffusionguard}.
In our evaluation, we measure this by comparing edits from purified images against the corresponding clean-image edit baseline.

The attacker may also fine-tune a personalization model (e.g., DreamBooth) on purified images for style mimicry.

The defender's goal is to ensure that the protection remains effective under attacker-side preprocessing and model mismatch: after purification and editing with a different target editor than the surrogate, the resulting edits should remain unreliable or substantially degraded relative to clean-image edits.
We formalize this setting and the resulting evaluation pipeline in Section~\ref{sec:setup}.

\section{Formulation}\label{sec:setup}

Building on the threat model in Section~\ref{sec:threat}, we formalize the deployment setting by expressing the attacker pipeline as a purification operator $\mathcal{P}$ followed by a target editor $\mathcal{E}$.
Crucially, the defender crafts protective perturbations against a surrogate editor, while the attacker may choose a different purification and a different editor.
We introduce notation and present an evaluation framework after release for measuring protection survivability under this model mismatch.

\subsection{Preliminary}\label{sec:preliminary}

\noindent\textbf{Notation.}
Let $x \in \mathbb{R}^{H \times W \times 3}$ denote a clean image.
A protection mechanism $\mathcal{D}$ produces a protected image $x_{\mathrm{adv}} = \mathcal{D}(x) = x + \delta$.

Here $\delta$ denotes an adversarial perturbation that satisfies $\|\delta\|_\infty \leq \epsilon$ for a fixed budget $\epsilon$.
A diffusion-based editing pipeline (editor) $\mathcal{E}$ takes an image and an editing instruction $y$.
The instruction may include text prompts and inpainting masks.
The editor outputs an edited image $\mathcal{E}(x, y)$.
A purification operator $\mathcal{P}$ is a preprocessing function that maps protected images toward the natural image manifold, producing $x_{\mathrm{pur}} = \mathcal{P}(x_{\mathrm{adv}})$.
When the attacker performs no explicit purification, $\mathcal{P}$ can be the identity operator.

\noindent\textbf{Problem Formulation.}
Under the threat model defined in Section~\ref{sec:threat}, the defender commits to a protection strategy $\mathcal{D}$ before release.
The attacker controls both the purification operator $\mathcal{P}$ and the target editor $\mathcal{E}$ after observing $x_{\mathrm{adv}}$.
The complete attack pipeline is defined as
\begin{equation}\label{eq:attack-pipeline}
x_{\mathrm{edit}} = \mathcal{E}(\mathcal{P}(x_{\mathrm{adv}}), y) = \mathcal{E}(x_{\mathrm{pur}}, y).
\end{equation}
The attacker succeeds if $x_{\mathrm{edit}}$ achieves editing quality comparable to the clean image baseline $x_{\mathrm{edit}}^{*} = \mathcal{E}(x, y)$, as measured in our evaluation; this baseline is not assumed available to the attacker.

\subsection{Key Insight}\label{sec:key-insight}

A fundamental difference distinguishes adversarial behavior in generative editing from that in classification. In classification, transfer failure means that an adversarial perturbation does not cause misclassification on a different model, and the attack simply fails. In generative editing, transfer failure can actively remove the protective signal. When $\mathcal{P}$, or even $\mathcal{E}$ itself, employs a generative prior that differs from the defender's surrogate, the reconstruction process projects $x_{\mathrm{adv}}$ back onto the natural image manifold, yielding a purified image editable without restriction.
This \emph{purify-once, edit-freely} vulnerability motivates a systematic study of protection survivability under purification after release.

\noindent\textbf{Threat Surface Taxonomy.}
To systematically vary attacker capability in this setting, we vary two attacker-controlled components: the target editor $\mathcal{E}$ and whether the attacker prepends a preprocessing operator $\mathcal{P}$ before editing.
This yields three primary settings that we use throughout the evaluation:
\ding{182}~\emph{Matched-surrogate}, where the target editor matches the surrogate model used to craft the perturbation.
\ding{183}~\emph{Editor mismatch}, where the target editor differs from the surrogate model (e.g., SD~v1.5 to SD~v2.0 or other diffusion-based editing pipelines).
\ding{184}~\emph{Preprocess then edit}, where the attacker first transforms $x_{\mathrm{adv}}$ using $\mathcal{P}$ and then edits the resulting image; we evaluate this setting with both matched and mismatched target editors to isolate the effect of $\mathcal{P}$ under model mismatch.

\subsection{Perturbation Purification Framework}\label{sec:framework}

To evaluate perturbation-based image protection under the threat model in Section~\ref{sec:threat}, we introduce a unified purification framework for the setting after release.
This framework targets the same post-release setting: after publishing $x_{\mathrm{adv}}$, the defender cannot observe or constrain the attacker-side pipeline, including the choice of purification operator $\mathcal{P}$ and downstream editor $\mathcal{E}$.
The framework explicitly captures attacker-controlled preprocessing prior to downstream semantic editing.
It is designed for realistic settings after release, where attackers independently choose both the purification strategy and the editing system from an open ecosystem.
No constraints are imposed by the defender.

\noindent\textbf{Evaluation Protocol.}
Evaluation compares the edited result $x_{\mathrm{edit}}$ with the clean image baseline $x_{\mathrm{edit}}^{*} = \mathcal{E}(x, y)$, which is used only for evaluation and is not assumed available to the attacker.
We use quantitative and perceptual metrics described in Section~\ref{sec:experiments}.
A protection survives a given attack pipeline if downstream edits remain unreliable or significantly degraded after purification, relative to clean image edits.
This formulation isolates the effect of purification.
It directly measures whether protective perturbations remain effective under reconstruction by heterogeneous attacker pipelines.

\noindent\textbf{Modular Design.}
The framework adopts a modular design.
The purification operator $\mathcal{P}$ can be instantiated as a simple image transformation, such as JPEG compression.
It can also take the form of an optimization based reconstruction method, such as IMPRESS~\cite{cao2023impress}.
This design enables reproducible and interpretable comparisons across protections, purifiers, and editors.
It directly tests the generality and practicality assumptions of the threat model.
The framework does not aim to construct the strongest adaptive attack.
Instead, it evaluates whether perturbation-based protections remain effective under practical attacker pipelines after release.
All purification operators act as lightweight preprocessing steps that do not rely on access to the defense internals.
They rely only on public models or data and require no access to protected images, gradients, or defense internals.
Failure under this framework indicates insufficient robustness to realistic model or distribution shift, rather than reliance on an overly powerful attacker.

\section{Methodology}\label{sec:method}

To instantiate the framework, we propose two purification methods, \textbf{VAE-Trans} in Section~\ref{sec:vae-trans} and \textbf{EditorClean} in Section~\ref{sec:editorclean}.
These methods are designed to probe distinct failure modes of perturbation-based image protection.
Our goal is to design two practical purifiers that isolate where protections break under the mismatch between the defender's surrogate editor and the attacker-side pipeline described in Section~\ref{sec:threat}.

Specifically, \textbf{VAE-Trans} isolates latent space distribution mismatch within the same model family (encoder mismatch), while \textbf{EditorClean} isolates reconstruction under a different generative prior (architecture mismatch).
Both methods satisfy two key design criteria: (i)~they require no protected images, gradients, or access to defense internals, and (ii)~they are practically realizable using only publicly available models and data.
Figure~\ref{fig:overview} illustrates the overall pipeline of the \textbf{EditorClean} purification method.

\subsection{\textbf{VAE-Trans}: Latent Space Purification}\label{sec:vae-trans}

\textbf{VAE-Trans} probes whether adversarial perturbations remain effective when images are processed by a VAE encoder with a shifted latent distribution, while keeping the decoder parameters fixed.
This setting commonly arises in practice due to encoder retraining, fine-tuning, or variant initialization within the same diffusion model family, such as Stable Diffusion v1.x variants.

\noindent\textbf{Formulation.}
Let $E$ and $D$ denote the encoder and decoder of a pretrained VAE.
A clean image $x$ is mapped to a latent representation $z = E(x)$ and reconstructed as $\hat{x} = D(z)$.
For a protected image $x_{\mathrm{adv}} = x + \delta$, the perturbation $\delta$ is crafted against a surrogate editing pipeline that uses the original encoder $E$.
\textbf{VAE-Trans} introduces a modified encoder $\tilde{E}$ that projects protected images back onto the natural image manifold:
\begin{equation}
x_{\mathrm{pur}} = D(\tilde{E}(x_{\mathrm{adv}})).
\end{equation}
The decoder $D$ remains frozen during both training and inference, ensuring that any degradation of protection arises from changes in latent encoding rather than decoder capacity.

\noindent\textbf{Training.}

To train $\tilde{E}$, we construct noisy clean image pairs from natural images $\{x_i\}_{i=1}^{N}$ by injecting Gaussian noise:
\begin{equation}
		\tilde{x}_i = x_i + \eta_i, \quad \eta_i \sim \mathcal{N}(0, \sigma^2 I),
		\end{equation}
where $\sigma = 0.1$. We choose this noise level to roughly match typical perturbation magnitudes (e.g., $\epsilon = 16/255$) while remaining visually mild, and we treat it as a design choice rather than a guarantee.
The training objective minimizes the mean squared error between the latent representation of the noisy image and that of the corresponding clean image:
\begin{equation}
\mathcal{L}_{\text{MSE}} = \frac{1}{d} \left\| \tilde{E}(\tilde{x}_i) - E(x_i) \right\|_2^2,
\end{equation}
where $d$ denotes the latent dimensionality and $E$ is the original frozen encoder.
This latent space alignment encourages $\tilde{E}$ to project noisy inputs back onto the clean latent manifold without involving the decoder during training.

\noindent\textbf{Interpretation.}
If protection fails under \textbf{VAE-Trans}, the adversarial perturbation is tightly coupled to a specific encoder distribution and does not transfer robustly even within the same model family.

This vulnerability can be exploited in practice by attackers who use an alternative VAE checkpoint or a retrained encoder variant.

\subsection{\textbf{EditorClean}: Instruction Guided Purification}\label{sec:editorclean}

\textbf{EditorClean} formulates image purification as an instruction guided semantic reconstruction task.
It leverages the strong generative priors of large scale Diffusion Transformers (DiTs).
Unlike \textbf{VAE-Trans}, which operates within the same architectural family as protected models, \textbf{EditorClean} deliberately adopts a heterogeneous DiT-based backbone rather than a UNet-based one.

This architectural heterogeneity is critical, as protective perturbations often exploit surrogate specific features and may not transfer to fundamentally different models.

\noindent\textbf{Formulation.}
\textbf{EditorClean} builds on the ICEdit in-context editing framework~\cite{zhang2025enabling}.
This framework enables instruction-based image editing through a diptych structure, where the source image appears on the left and the generated output appears on the right. We view purification as a special case of editing: the desired transformation preserves the scene while removing small protective perturbations. This perspective makes ICEdit's in-context learning paradigm a natural fit for the purification objective.
We adapt this paradigm for purification by conditioning on a fixed denoising instruction $y_{\mathrm{denoise}}$:
\begin{equation}
x_{\mathrm{pur}} = \mathcal{G}_\theta(x_{\mathrm{adv}},\, y_{\mathrm{denoise}}),
\end{equation}
where $\mathcal{G}_\theta$ denotes a DiT based generator that reconstructs a clean image by treating the protected input as a semantic reference.
We adopt FLUX.1-fill-dev~\cite{labs2025flux} as the backbone, a 12B parameter DiT based inpainting model.
Following the parameter efficient adaptation strategy of ICEdit~\cite{zhang2025enabling}, we apply LoRA fine-tuning~\cite{hu2022lora}, introducing approximately 0.1\% additional trainable parameters relative to the base model.

\noindent\textbf{Training.}

Following the ICEdit protocol, we construct diptych training pairs from natural images $\{x_i\}_{i=1}^{N}$ by injecting Gaussian noise:
\begin{equation}
\tilde{x}_i = x_i + \eta_i, \quad \eta_i \sim \mathcal{N}(0, \sigma_{\mathrm{train}}^2 I),
\end{equation}
where $\sigma_{\mathrm{train}} = 0.1$. We train with a fixed noise level to encourage robustness to small, structured perturbations.
Each diptych places the noisy image $\tilde{x}_i$ on the left panel and the clean target image $x_i$ on the right panel.
We employ the ICEdit in-context prompt template to define the denoising instruction $y_{\mathrm{denoise}}$:
\emph{``A diptych with two side by side images of the same scene. On the right, the scene is exactly the same as on the left but with the noise removed.''}
The model is trained to reconstruct $x_i$ from $\tilde{x}_i$ conditioned on $y_{\mathrm{denoise}}$ by minimizing the reconstruction loss:

\begin{equation}
\mathcal{L}_{\text{edit}} = \mathbb{E}_{x_i,\eta_i}\left[\left\|\mathcal{G}_\theta(\tilde{x}_i,\, y_{\mathrm{denoise}}) - x_i\right\|_2^2\right].
\end{equation}
Here $\mathbb{E}$ denotes expectation over the sampled pairs $(x_i,\eta_i)$.
This process leverages the in-context learning capability of DiT to establish strong correspondence between the reference and purified panels.

\noindent\textbf{Inference.}
At inference time, we inject additional Gaussian noise to disrupt structured adversarial patterns before purification:
\begin{equation}
x_{\mathrm{pur}} = \mathcal{G}_\theta(x_{\mathrm{adv}} + \eta,\, y_{\mathrm{denoise}}), \quad \eta \sim \mathcal{N}(0, \sigma_{\mathrm{test}}^2 I),
\end{equation}
with $\sigma_{\mathrm{test}} = 0.10$, which we validate through the ablation studies in Section~\ref{sec:ablation}.
This stochastic perturbation weakens finely optimized spatial correlations in the adversarial signal.
The instruction guided generation then reconstructs a semantically aligned output.

\noindent\textbf{Interpretation.}
The effectiveness of \textbf{EditorClean} arises from two complementary factors.
First, \textbf{model heterogeneity}: the DiT backbone differs substantially from the UNet architectures that many protections are optimized against; as a result, perturbations crafted for a UNet-based surrogate often transfer poorly to a DiT-based purifier. Importantly, this mismatch can also provide purification even when the protection is crafted on a DiT editor: perturbations optimized for one DiT model may still transfer poorly to another (e.g., Step1X-Edit~\cite{liu2025step1x} vs.\ FLUX.1-fill-dev~\cite{labs2025flux}), so switching DiT backbones can itself attenuate the protective signal (see Section~\ref{sec:step1x}).
Second, \textbf{semantic reconstruction}: the generative prior emphasizes semantic consistency over low-level pixel correlations, which encourages the model to suppress fine-grained adversarial textures during reconstruction.
Section~\ref{sec:experiments} evaluates how well this approach restores editability under model mismatch.

\section{Experiments}\label{sec:experiments}

We conduct comprehensive experiments to evaluate the robustness of perturbation-based image protections under realistic deployment conditions with model mismatch.

To systematically study model mismatch, we use a multi-stage evaluation protocol that progressively increases the discrepancy between the defender's surrogate editor and the attacker-side pipeline. We start with a \emph{matched-surrogate baseline}, where both protection and evaluation use SD~v1.5. We then consider a \emph{cross-version transfer} setting, evaluating perturbations crafted on SD~v1.5 against SD~v2.0. Finally, we evaluate the most realistic setting of \emph{purify then edit}, where the attacker prepends a purification operator before downstream editing.
Through this unified evaluation framework, we aim to answer three research questions:

\ding{182}~To what extent do existing adversarial protections remain effective beyond the matched-surrogate setting for which they are optimized?
\ding{183}~How does increasing model mismatch, including cross version editors, purification preprocessing, and architecturally distinct models, affect the robustness of protective perturbations?
\ding{184}~How do different purification strategies compare in their ability to remove protective perturbations and restore editability under realistic deployment conditions?

\subsection{Experimental Setup}

\noindent\textbf{Targeted Models.}
We evaluate adversarial protections on diffusion-based image editing pipelines from the Stable Diffusion family.

In our main SD benchmark, all protective perturbations are optimized against SD~v1.5 Inpainting as the matched-surrogate target model.
SD v1.5 Inpainting serves as the matched-surrogate target because it is the most widely used editing backbone and the primary target of existing protection methods. SD v2.0 Inpainting provides a natural stress test for cross-version transferability.

\noindent\textbf{Dataset.}
We conduct all evaluations on the DiffusionGuard dataset~\cite{choi2024diffusionguard}, comprising 2,100 text guided inpainting tasks.
We follow the original evaluation protocol to ensure fair comparison across methods.
See Appendix~\ref{sec:dataset_details} for details.

\noindent\textbf{Protection Methods.}
We evaluate six representative perturbation-based protection methods spanning different attack strategies and optimization objectives:
\ding{182}~\textbf{PhotoGuard}~\cite{salman2023raising}: Optimizes perturbations against the VAE encoder to distort latent representations toward a target image.
\ding{183}~\textbf{AdvDM}~\cite{liang2023adversarial}: Applies adversarial perturbations optimized against the diffusion denoising process.
\ding{184}~\textbf{MIST}~\cite{liang2023mist}: Incorporates texture aware losses to prevent unauthorized style transfer.
\ding{185}~\textbf{SDS}~\cite{DBLP:conf/iclr/XueLWC24}: Leverages score distillation sampling for efficient gradient approximation.
\ding{186}~\textbf{DiffusionGuard}~\cite{choi2024diffusionguard}: Targets text-image alignment during the early denoising stages with mask augmentation.
\ding{187}~\textbf{AdvPaint}~\cite{jeon2025advpaintprotectingimagesinpainting}: Optimizes perturbations specifically for inpainting based editing scenarios.
All protected images are generated under a fixed perturbation budget of $\epsilon = 16/255$ in the $\ell_\infty$ norm.
In our main SD benchmark, all protective perturbations are trained against SD~v1.5 Inpainting only, which serves as the surrogate model; evaluations on SD~v2.0 and other pipelines therefore reflect strictly cross-model transfer scenarios.
In Section~\ref{sec:step1x}, we additionally adapt three protection methods to support DiT-based perturbation generation and re-optimize them on the DiT editor Step1X-Edit.

\noindent\textbf{Purification Methods.}
We instantiate the purification operator $\mathcal{P}$ with both proposed and baseline methods.
\emph{Proposed methods:}

\ding{182}~\textbf{VAE-Trans} (Section~\ref{sec:vae-trans}): A VAE encoder adapted via fine-tuning that maps protected images to corrected latent representations. For each target editor, we train a dedicated \textbf{VAE-Trans} model by fine-tuning the matching VAE encoder while keeping the decoder frozen, using the SD~v1.5 VAE for SD~v1.5 Inpainting and the SD~v2.0 VAE for SD~v2.0 Inpainting.
\ding{183}~\textbf{EditorClean} (Section~\ref{sec:editorclean}): An instruction guided purifier, using FLUX.1-fill-dev~\cite{labs2025flux} as backbone with LoRA based fine-tuning for denoising.
\emph{Baselines:}
\ding{184}~\textbf{IMPRESS}~\cite{cao2023impress}: An optimization based consistency restoration method that iteratively refines protected images to satisfy self-consistency constraints under VAE reconstruction.
\ding{185}~\textbf{GridPure}~\cite{zhao2024can}: A patch level reconstruction baseline that processes overlapping image patches independently and aggregates the denoised outputs.
\ding{186}~\textbf{JPEG compression}~\cite{wallace1991jpeg,sandoval2023jpeg}: A simple preprocessing baseline that attenuates high frequency adversarial perturbations through lossy compression. Together, these baselines span distinct purification paradigms established in prior work~\cite{honig2024adversarial,zhao2024can}, ranging from lightweight signal processing (JPEG) through iterative optimization (IMPRESS) to learned generative reconstruction (GridPure), enabling comprehensive comparison across a broad range of purification strategies.

\noindent\textbf{Evaluation Metrics.}
We assess performance using four metrics:
\ding{182}~\textbf{Peak Signal-to-Noise Ratio (PSNR)}: Measures pixel level similarity to a paired clean reference. Higher PSNR indicates better reconstruction fidelity.
\ding{183}~\textbf{Learned Perceptual Image Patch Similarity (LPIPS)}~\cite{zhang2018unreasonable}: Evaluates perceptual similarity to a paired clean reference using deep features from pretrained networks. Lower LPIPS indicates better perceptual fidelity.
\ding{184}~\textbf{Fr\'{e}chet Inception Distance (FID)}~\cite{heusel2017gans}: Measures distributional similarity between a set of outputs and the corresponding clean reference set. Lower FID indicates that purification does not introduce artifacts or distribution shifts.
\ding{185}~\textbf{ImageReward (IR)}~\cite{xu2023imagereward}: A learned prompt-conditioned reward model trained from human preference comparisons. Higher IR is typically associated with fewer visible artifacts, stronger prompt-image alignment, and more human-preferred aesthetics (e.g., fewer distorted limbs or incorrect numbers of body parts). Notably, IR does not directly measure object/identity consistency with the original image in editing.
Unless stated otherwise, for downstream editing we compute PSNR/LPIPS/FID between the edited outputs produced from protected inputs (optionally purified) and the edited outputs produced from the corresponding clean images, $x_{\mathrm{edit}}^{*}=\mathcal{E}(x,y)$. For purified-image quality (before editing), we compute LPIPS/FID between purified outputs and the corresponding clean images $x$.
PSNR and LPIPS are computed per instance and averaged over the dataset, while FID is computed once per setting between the output set and the corresponding clean reference set.

\subsection{Implementation Details}
Unless stated otherwise, all reported training and inference times are measured on a single NVIDIA RTX 3090 GPU.

\noindent\textbf{Defense Generation.}
All adversarially protected images are generated using the official implementations of each protection method with their default hyperparameters. For optimization based defenses, we follow the standard DiffusionGuard protocol~\cite{choi2024diffusionguard} and optimize perturbations using projected gradient descent (PGD)~\cite{madry2017towards} with 100 iterations, step size $\alpha = 2/255$, and a fixed $\ell_\infty$ budget of $\epsilon = 16/255$. This setup matches the configurations adopted in prior protection works~\cite{salman2023raising,liang2023adversarial,liang2023mist,choi2024diffusionguard,jeon2025advpaintprotectingimagesinpainting} and recent robustness studies~\cite{honig2024adversarial,zhao2024can}, ensuring consistency with existing evaluation protocols and comparability across defenses.

For the SD benchmark experiments in this section, all protective perturbations are optimized exclusively against SD~v1.5 Inpainting, which serves as the matched-surrogate target model.

\noindent\textbf{Purification Training.}
Both purification methods are trained on a subset of the \textbf{OmniEdit-Filtered-1.2M} dataset~\cite{wei2024omniedit}. We sample about 2,000 images and inject synthetic Gaussian noise ($\sigma = 0.1$) to simulate adversarial perturbations with magnitude comparable to typical $\ell_\infty$ budgets ($\approx 16/255$).
For \textbf{VAE-Trans}, we fine-tune only the VAE encoder while keeping the decoder frozen to preserve the pretrained generative prior.

We train \textbf{VAE-Trans} using the VAE corresponding to the target editor: for SD~v1.5 Inpainting we fine-tune the SD~v1.5 VAE encoder, and for SD~v2.0 Inpainting we fine-tune the SD~v2.0 VAE encoder.
Training uses the AdamW optimizer~\cite{loshchilov2017decoupled} with learning rate $1 \times 10^{-4}$ for 6 epochs and takes about 1 hour.
For \textbf{EditorClean}, we adapt the FLUX.1-fill-dev~\cite{labs2025flux} model via LoRA fine-tuning using the denoising prompt described in Section~\ref{sec:editorclean}. We train with batch size 2 for 2,000 steps, which takes about 4 hours.

\noindent\textbf{Inference and Editing.}

At \textbf{EditorClean} test time, we apply light Gaussian noise prior to purification to attenuate high-frequency perturbation components ($\sigma_{\mathrm{test}} = 0.10$, unless stated otherwise).
For purification, no information about the defense algorithm, perturbation generation process, or gradients is assumed available to the attacker.
In terms of computational efficiency, \textbf{EditorClean} processes each image in approximately 1 minute, compared with 3 minutes for GridPure and 12 minutes for IMPRESS. \textbf{VAE-Trans} can be applied by swapping the VAE encoder and does not introduce additional iterative inference steps.
All downstream editing experiments are performed using 50 DDIM~\cite{song2020denoising} sampling steps with classifier-free guidance~\cite{ho2022classifier} scale of 7.5, following the DiffusionGuard evaluation pipeline~\cite{choi2024diffusionguard}, where a fixed random seed is used to limit sampling randomness in downstream editing.

\begin{table}[t]
\centering
\footnotesize
\renewcommand{\arraystretch}{0.9}

\caption{Purified image quality compared to clean images.
	We report LPIPS$\downarrow$ and FID$\downarrow$ between purified images and their clean counterparts.}
\label{tab:purified_quality}
\begin{tabular}{llcc}
\hline
\textbf{Protection} & \textbf{Purifier} & \textbf{LPIPS$\downarrow$} & \textbf{FID$\downarrow$} \\
\hline
\multirow{5}{*}{AdvDM}
 & Unpurified    & 0.458 & 165.58 \\
 & JPEG        & 0.456 & 160.69 \\
 & IMPRESS     & 0.533 & 234.10 \\
 & GridPure    & 0.337 & 144.03 \\
 & \textbf{EditorClean} (ours) & \textbf{0.244} & \textbf{91.29} \\
\hline
\multirow{5}{*}{AdvPaint}
 & Unpurified    & 0.338 & 80.49 \\
 & JPEG        & 0.317 & 81.58 \\
 & IMPRESS     & 0.540 & 241.29 \\
 & GridPure    & 0.137 & 67.37 \\
 & \textbf{EditorClean} (ours) & \textbf{0.130} & \textbf{54.17} \\
\hline
\multirow{5}{*}{DiffusionGuard}
 & Unpurified    & 0.258 & 51.65 \\
 & JPEG        & 0.232 & 52.68 \\
 & IMPRESS     & 0.418 & 134.63 \\
 & GridPure    & 0.117 & 63.60 \\
 & \textbf{EditorClean} (ours) & \textbf{0.106} & \textbf{44.12} \\
\hline
\multirow{5}{*}{MIST}
 & Unpurified    & 0.460 & 121.19 \\
 & JPEG        & 0.450 & 121.49 \\
 & IMPRESS     & 0.579 & 290.37 \\
 & GridPure    & \textbf{0.250} & \textbf{102.16} \\
 & \textbf{EditorClean} (ours) & 0.297 & 108.46 \\
\hline
\multirow{5}{*}{PhotoGuard}
 & Unpurified    & 0.462 & 127.92 \\
 & JPEG        & 0.453 & 126.16 \\
 & IMPRESS     & 0.558 & 266.98 \\
 & GridPure    & \textbf{0.254} & \textbf{103.02} \\
 & \textbf{EditorClean} (ours) & 0.297 & 115.89 \\
\hline
\multirow{5}{*}{SDS}
 & Unpurified    & 0.393 & 93.72 \\
 & JPEG        & 0.403 & 97.79 \\
 & IMPRESS     & 0.479 & 152.96 \\
 & GridPure    & 0.141 & 64.69 \\
 & \textbf{EditorClean} (ours) & \textbf{0.135} & \textbf{60.33} \\
\hline
\end{tabular}
\end{table}

\begin{figure}[t]
\centering
\includegraphics[width=\columnwidth]{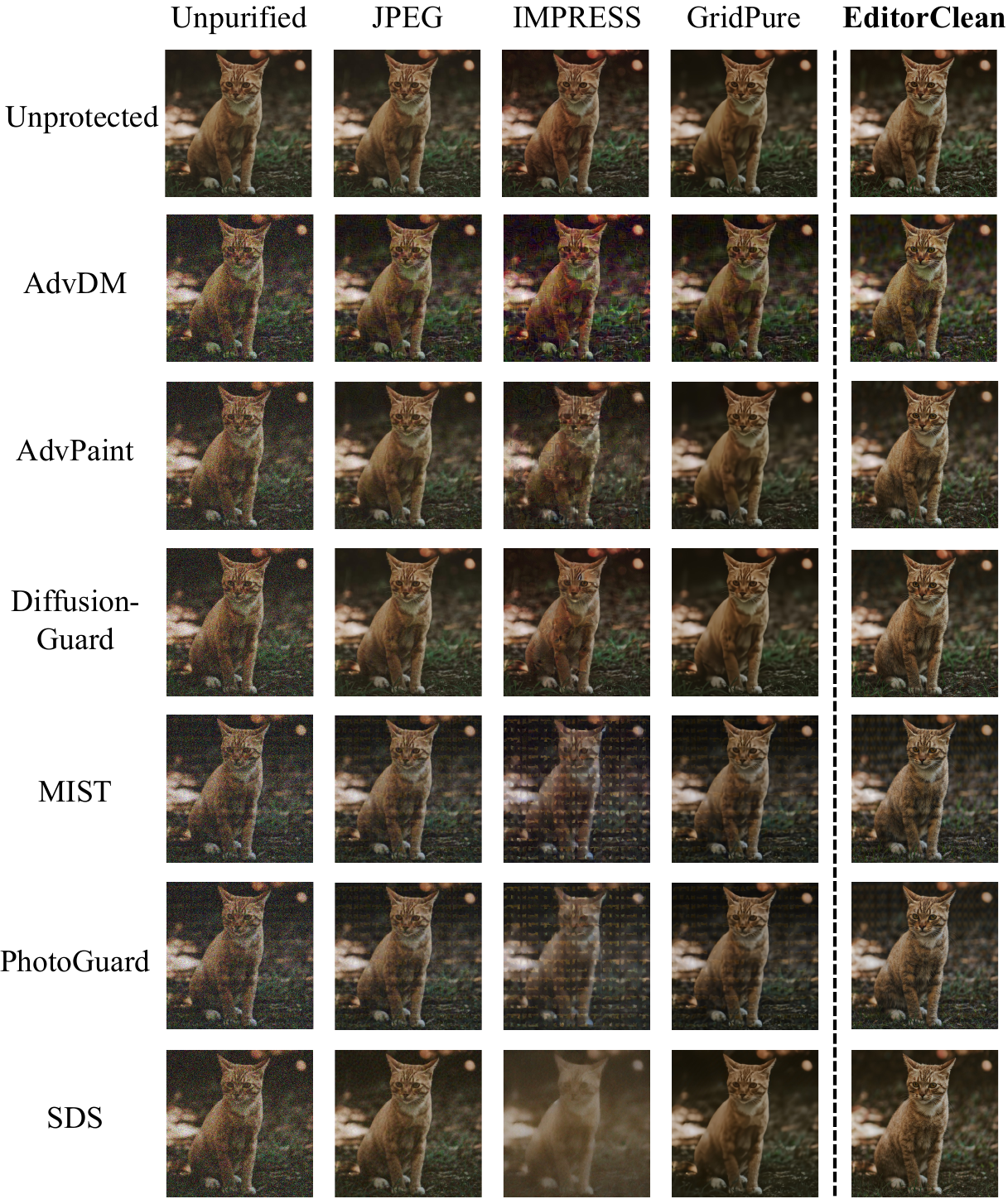}
\caption{Qualitative comparison of purified image quality across different protection methods and purification strategies (see Table~\ref{tab:purified_quality}). Each row shows results for one protection method.}
\label{fig:purified_quality_visual}
\end{figure}

\begin{table*}[t]
\centering
\footnotesize
\setlength{\tabcolsep}{10pt}
\renewcommand{\arraystretch}{0.83}
\caption{Quantitative comparison of downstream editing results obtained from purified images on SD~v1.5 Inpainting (left) and SD~v2.0 Inpainting (right, cross-version transfer). All protections are optimized against SD~v1.5 only. We report PSNR$\uparrow$, LPIPS$\downarrow$, FID$\downarrow$, and IR$\uparrow$. Best results within each protection are highlighted in bold. The last block reports the effect of applying each purification method to clean images.}
\label{tab:edited_result_comparison}
\begin{tabular}{ll|cccc|cccc}
\toprule
& & \multicolumn{4}{c|}{\textbf{SD~v1.5 Inpainting}} & \multicolumn{4}{c}{\textbf{SD~v2.0 Inpainting}} \\
\cmidrule(lr){3-6} \cmidrule(lr){7-10}
\textbf{Protection} & \textbf{Purifier} & \textbf{PSNR$\uparrow$} & \textbf{LPIPS$\downarrow$} & \textbf{FID$\downarrow$} & \textbf{IR$\uparrow$} & \textbf{PSNR$\uparrow$} & \textbf{LPIPS$\downarrow$} & \textbf{FID$\downarrow$} & \textbf{IR$\uparrow$} \\
\midrule
\multirow{6}{*}{AdvDM}
 & Unpurified    & 15.70 & 0.488 & 60.42 & 0.031 & 15.09 & 0.507 & 56.13 & 0.091 \\
 & JPEG        & 16.27 & 0.454 & 49.82 & 0.052 & 15.55 & 0.475 & 48.26 & 0.127 \\
 & IMPRESS     & 15.25 & 0.500 & 69.17 & -0.018 & 14.83 & 0.507 & 62.59 & 0.075 \\
 & GridPure    & 16.41 & 0.424 & 41.92 & 0.022 & 16.12 & 0.437 & 40.73 & 0.009 \\
 & \textbf{VAE-Trans} (ours)   & 16.38 & 0.438 & 39.64 & 0.044 & 15.90 & 0.445 & 37.79 & 0.133 \\
 & \textbf{EditorClean} (ours) & \textbf{18.53} & \textbf{0.310} & \textbf{24.53} & \textbf{0.129} & \textbf{18.31} & \textbf{0.301} & \textbf{23.13} & \textbf{0.211} \\
\midrule
\multirow{6}{*}{AdvPaint}
 & Unpurified    & 13.83 & 0.581 & 78.75 & -0.213 & 13.87 & 0.561 & 67.42 & -0.045 \\
 & JPEG        & 16.94 & 0.406 & 37.62 & \textbf{0.086} & 16.52 & 0.411 & 36.75 & 0.172 \\
 & IMPRESS     & 13.74 & 0.593 & 89.40 & -0.233 & 13.71 & 0.570 & 75.23 & -0.105 \\
 & GridPure    & 16.47 & 0.424 & 36.35 & -0.134 & 16.24 & 0.433 & 36.89 & -0.143 \\
 & \textbf{VAE-Trans} (ours)   & 17.30 & 0.374 & 27.93 & 0.018 & 17.04 & 0.368 & 27.23 & 0.131 \\
 & \textbf{EditorClean} (ours) & \textbf{19.35} & \textbf{0.270} & \textbf{21.07} & 0.054 & \textbf{18.97} & \textbf{0.269} & \textbf{20.06} & \textbf{0.190} \\
\midrule
\multirow{6}{*}{DiffusionGuard}
 & Unpurified    & 15.24 & 0.557 & 53.47 & -0.455 & 15.32 & 0.506 & 45.81 & -0.156 \\
 & JPEG        & 18.35 & 0.339 & 28.69 & 0.026 & 18.00 & 0.334 & 27.76 & 0.173 \\
 & IMPRESS     & 15.94 & 0.477 & 45.36 & -0.179 & 15.99 & 0.450 & 39.20 & -0.011 \\
 & GridPure    & 16.58 & 0.418 & 37.43 & -0.182 & 16.37 & 0.425 & 38.66 & -0.195 \\
 & \textbf{VAE-Trans} (ours)   & 17.66 & 0.354 & 25.94 & -0.000 & 17.46 & 0.345 & 24.73 & 0.147 \\
 & \textbf{EditorClean} (ours) & \textbf{19.32} & \textbf{0.271} & \textbf{19.89} & \textbf{0.054} & \textbf{19.12} & \textbf{0.260} & \textbf{19.22} & \textbf{0.223} \\
\midrule
\multirow{6}{*}{MIST}
 & Unpurified    & 14.57 & 0.555 & 71.83 & -0.244 & 14.18 & 0.552 & 68.04 & -0.134 \\
 & JPEG        & 16.32 & 0.463 & 47.24 & -0.040 & 15.77 & 0.462 & 43.18 & 0.141 \\
 & IMPRESS     & 14.10 & 0.582 & 94.21 & -0.385 & 13.70 & 0.581 & 87.84 & -0.298 \\
 & GridPure    & 16.08 & 0.442 & 38.37 & -0.119 & 15.82 & 0.448 & 38.54 & -0.115 \\
 & \textbf{VAE-Trans} (ours)   & 16.84 & 0.399 & 33.64 & 0.025 & 16.49 & 0.392 & 30.95 & \textbf{0.191} \\
 & \textbf{EditorClean} (ours) & \textbf{17.87} & \textbf{0.343} & \textbf{29.87} & \textbf{0.048} & \textbf{17.52} & \textbf{0.343} & \textbf{28.15} & 0.139 \\
\midrule
\multirow{6}{*}{PhotoGuard}
 & Unpurified    & 14.66 & 0.554 & 69.90 & -0.248 & 14.20 & 0.553 & 67.46 & -0.130 \\
 & JPEG        & 16.39 & 0.462 & 47.40 & -0.043 & 15.76 & 0.465 & 44.09 & 0.132 \\
 & IMPRESS     & 14.33 & 0.578 & 82.30 & -0.411 & 13.91 & 0.579 & 78.85 & -0.297 \\
 & GridPure    & 16.11 & 0.438 & 38.72 & -0.125 & 15.85 & 0.449 & 39.04 & -0.120 \\
 & \textbf{VAE-Trans} (ours)   & 16.88 & 0.398 & \textbf{33.72} & \textbf{0.021} & 16.56 & 0.389 & \textbf{31.07} & \textbf{0.180} \\
 & \textbf{EditorClean} (ours) & \textbf{17.52} & \textbf{0.365} & 33.87 & 0.015 & \textbf{17.16} & \textbf{0.368} & 31.31 & 0.095 \\
\midrule
\multirow{6}{*}{SDS}
 & Unpurified    & 16.03 & 0.523 & 42.55 & -0.033 & 15.66 & 0.494 & 40.64 & 0.077 \\
 & JPEG        & 16.98 & 0.459 & 38.85 & 0.036 & 16.53 & 0.448 & 35.27 & 0.141 \\
 & IMPRESS     & 15.46 & 0.564 & 52.15 & -0.308 & 15.01 & 0.554 & 50.53 & -0.246 \\
 & GridPure    & 16.51 & 0.416 & 34.11 & -0.113 & 16.30 & 0.424 & 34.49 & -0.106 \\
 & \textbf{VAE-Trans} (ours)   & 16.99 & 0.411 & 30.58 & 0.047 & 16.54 & 0.404 & 29.68 & 0.149 \\
 & \textbf{EditorClean} (ours) & \textbf{19.14} & \textbf{0.278} & \textbf{22.23} & \textbf{0.061} & \textbf{18.88} & \textbf{0.272} & \textbf{20.83} & \textbf{0.219} \\
\midrule
\multirow{6}{*}{Unprotected}
 & Unpurified   & - & - & - & 0.107 & - & - & - & 0.223 \\
 & JPEG        & 22.23 & 0.176 & 14.38 & 0.104 & 21.57 & 0.190 & 14.84 & 0.229 \\
 & IMPRESS     & 20.79 & 0.218 & 16.87 & 0.124 & 20.36 & 0.216 & 16.36 & 0.207 \\
 & GridPure    & 16.61 & 0.416 & 38.27 & -0.167 & 16.50 & 0.419 & 38.91 & -0.181 \\
 & \textbf{VAE-Trans} (ours)   & 17.55 & 0.369 & 26.69 & -0.038 & 17.33 & 0.366 & 26.93 & 0.057 \\
 & \textbf{EditorClean} (ours) & 19.58 & 0.258 & 19.30 & 0.063 & 19.28 & 0.253 & 18.65 & 0.219 \\
\bottomrule
\end{tabular}
\end{table*}

\begin{figure*}[t]
\centering
\includegraphics[width=\textwidth]{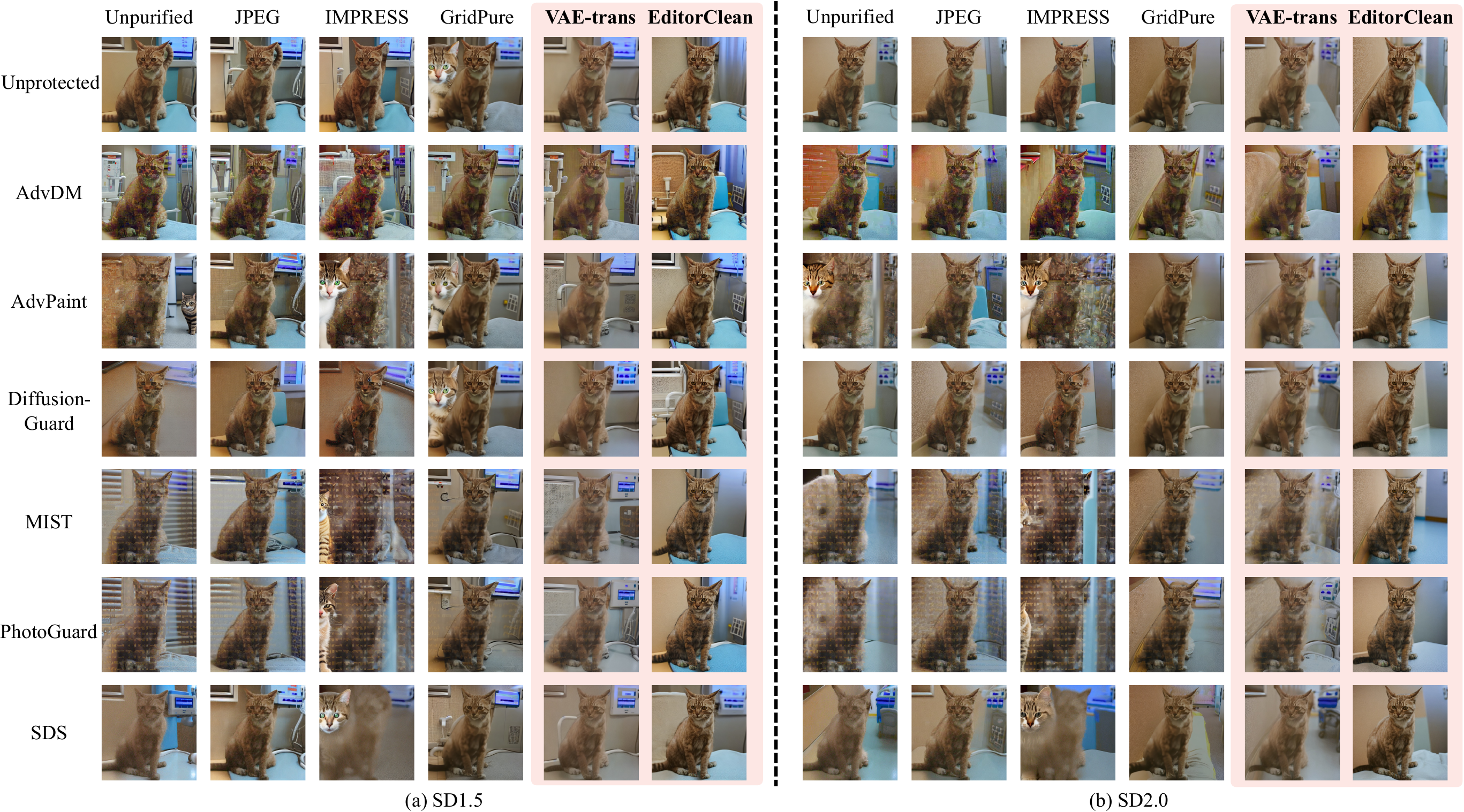}
\caption{Qualitative downstream editing results on SD~v1.5 Inpainting (left) and SD~v2.0 Inpainting (right). Rows correspond to six protection methods, and columns show edits produced from unpurified protected inputs or from protected inputs after different purification strategies (see Table~\ref{tab:edited_result_comparison}).}
\label{fig:sd15_sd20_gen}
\end{figure*}

\subsection{Reconstruction Quality of Purified Images}

Table~\ref{tab:purified_quality} summarizes the visual fidelity of purified images. Even without purification, protected images exhibit significant distortions: average LPIPS exceeds 0.38 and FID ranges from 51.65 to 165.58, suggesting the protection strength is at the upper limit of realistic deployment. Specifically, PhotoGuard and MIST introduce the strongest distortions (LPIPS 0.458), while DiffusionGuard maintains better integrity (LPIPS 0.258).
Among purification methods, GridPure achieves competitive LPIPS and FID on PhotoGuard and MIST via aggressive patch-scale smoothing. However, this pixel-level restoration does not translate into superior editability. As shown in Table~\ref{tab:edited_result_comparison}, \textbf{EditorClean} consistently outperforms GridPure in downstream tasks, improving PSNR from 16.11 to 17.52 on PhotoGuard and from 16.08 to 17.87 on MIST. This indicates that effective purification hinges on neutralizing adversarial signals rather than merely restoring pixel similarity.
\textbf{EditorClean} strikes the optimal balance between fidelity and robustness: it achieves the best purification quality across the majority of protection methods (e.g., AdvDM, AdvPaint, DiffusionGuard, and SDS), while simultaneously delivering peak downstream editing performance. For instance, on DiffusionGuard, it significantly reduces LPIPS to 0.106 and FID to 44.12, outperforming all baselines.
In contrast, IMPRESS consistently degrades image quality, yielding substantially higher FID than the unpurified baselines. We attribute this to two factors: (1) excessive noise during diffusion reconstruction which destroys fine details, and (2) error accumulation across iterative VAE encoding/decoding cycles.
Qualitative results in Figure~\ref{fig:purified_quality_visual} align with these metrics. While IMPRESS suffers from over-smoothing and GridPure occasionally retains subtle adversarial patterns, \textbf{EditorClean} consistently produces high-fidelity images that effectively neutralize protective perturbations.

\subsection{Purification Restores Editability}

Table~\ref{tab:edited_result_comparison} evaluates purification effectiveness through downstream editing quality on SD~v1.5 and SD~v2.0.
This evaluation directly measures whether adversarial protections remain effective after preprocessing.
Since all protections are optimized only for SD~v1.5, results on SD~v2.0 provide a strict test of cross version transferability.

\noindent\textbf{Without Purification.}
All protection methods substantially degrade editing quality on both editors.

PSNR remains around 14 to 16, and LPIPS is typically around 0.49 to 0.58.
AdvPaint causes the strongest disruption and reaches a PSNR of 13.83 on SD~v1.5 and 13.87 on SD~v2.0.
However, protection does not transfer perfectly across versions.
Unpurified images edited by SD~v2.0 can be slightly more editable than those edited by SD~v1.5, depending on the protection method.
For example, DiffusionGuard achieves a PSNR of 15.32 on SD~v2.0, compared with 15.24 on SD~v1.5.
This gap indicates an inherent cross model weakness even without explicit purification.

\noindent\textbf{With Purification.}

\textbf{EditorClean} consistently achieves the strongest restoration of editing quality across all six protection methods and both editors. Compared with unpurified inputs, it improves PSNR by 3–5 dB and reduces FID by approximately 50–70\%, bringing purified edits close to clean-image baselines in both reconstruction accuracy and perceptual quality.
For instance, on AdvPaint protection, \textbf{EditorClean} raises PSNR from 13.83 to 19.35 (a 40\% improvement) and reduces FID from 78.75 to 21.07 (a 73\% reduction), bringing editing quality close to the clean-image baseline (PSNR 19.58, FID 19.30).
We attribute \textbf{EditorClean}'s advantage over GridPure to two main factors.
The first factor is architectural mismatch.
GridPure relies on the same SD family model as the target editors, while \textbf{EditorClean} adopts a DiT based architecture that protective perturbations do not target.
The second factor is in-context learning.
By leveraging the ICEdit paradigm~\cite{zhang2025enabling} with task specific denoising instructions, \textbf{EditorClean} suppresses adversarial artifacts more precisely than the generic patch level reconstruction used by GridPure.
To complement similarity-based fidelity metrics, Table~\ref{tab:edited_result_comparison} also reports ImageReward (IR)~\cite{xu2023imagereward}, which reflects human-preferred visual quality and prompt-image alignment of the edited outputs.
IR is mainly influenced by visible artifacts, prompt adherence, and perceived aesthetics (e.g., distorted limbs or incorrect numbers of fingers/arms/tails).
Across most defenses, purification improves IR relative to unpurified inputs, consistent with fewer artifacts and stronger prompt following.
We note that IR does not explicitly evaluate object/identity consistency with the input image, which helps explain why the highest IR does not always coincide with the strongest edit preservation.

These gains remain stable under cross version transfer.
On AdvPaint, \textbf{EditorClean} consistently improves editing quality across different SD versions.
This result indicates that purification performance is largely independent of the target editor.
\textbf{VAE-Trans} achieves competitive performance with substantially lower computational cost.
It delivers the best or near best overall quality on PhotoGuard.
Even simple baselines such as JPEG compression~\cite{wallace1991jpeg,sandoval2023jpeg} partially restore editability.
This observation shows that current protections remain vulnerable to basic image transformations.
In contrast, IMPRESS performs poorly overall and in some settings underperforms the no purification baseline.
This suggests that optimization based reconstruction can introduce distortions that hinder downstream editing.
Overall, these results show that existing perturbation-based protections exhibit systematic vulnerability to purification after release under model mismatch.
This vulnerability persists across different editor versions and does not require any prior knowledge of the target model.

\begin{table}[t]
\centering
\footnotesize
\setlength{\tabcolsep}{6pt}
\renewcommand{\arraystretch}{0.9}
\caption{Ablation study on Gaussian noise strength $\sigma$ before \textbf{EditorClean} purification.
We report PSNR$\uparrow$, LPIPS$\downarrow$, FID$\downarrow$, and IR$\uparrow$ between edited results of purified protected images and those edited from clean images.}
\label{tab:gaussian_ablation}
\begin{tabular}{lccccc}
\toprule
\textbf{Method} & $\boldsymbol{\sigma}$ & \textbf{PSNR} $\uparrow$ & \textbf{LPIPS} $\downarrow$ & \textbf{FID} $\downarrow$ & \textbf{IR} $\uparrow$ \\
\midrule
\multirow{4}{*}{AdvDM}
 & 0.05 & 17.55 & 0.3648 & 33.23 & 0.1127 \\
 & 0.10 & \textbf{18.53} & \textbf{0.3095} & \textbf{24.53} & 0.1294 \\
 & 0.15 & 18.43 & 0.3112 & 24.57 & \textbf{0.1623} \\
 & 0.20 & 18.23 & 0.3222 & 25.77 & 0.1527 \\
\midrule
\multirow{4}{*}{AdvPaint}
 & 0.05 & \textbf{19.36} & 0.2726 & \textbf{20.04} & 0.0689 \\
 & 0.10 & 19.35 & \textbf{0.2696} & 21.07 & 0.0535 \\
 & 0.15 & 19.11 & 0.2775 & 21.82 & 0.0889\\
 & 0.20 & 18.43 & 0.3060 & 23.85 & \textbf{0.0989}  \\
\midrule
\multirow{4}{*}{DiffusionGuard}
 & 0.05 & \textbf{19.55} & \textbf{0.2622} & \textbf{19.56} & 0.0847 \\
 & 0.10 & 19.32 & 0.2709 & 19.89 & 0.0538 \\
 & 0.15 & 18.87 & 0.2861 & 21.83 & 0.0886 \\
 & 0.20 & 18.59 & 0.3017 & 23.11 & \textbf{0.1055} \\
\midrule
\multirow{4}{*}{MIST}
 & 0.05 & 16.83 & 0.3970 & 39.26 & 0.0339 \\
 & 0.10 & 17.87 & 0.3432 & 29.87 & 0.0481 \\
 & 0.15 & 18.30 & \textbf{0.3195} & 26.04 & 0.0369 \\
 & 0.20 & \textbf{18.23} & 0.3207 & \textbf{24.98} & \textbf{0.1084} \\
\midrule
\multirow{4}{*}{PhotoGuard}
 & 0.05 & 16.79 & 0.4002 & 40.61 & 0.0241 \\
 & 0.10 & 17.52 & 0.3646 & 33.87 & 0.0150 \\
 & 0.15 & \textbf{18.31} & \textbf{0.3223} & \textbf{26.03} & 0.0772 \\
 & 0.20 & 18.19 & 0.3260 & 26.56 & \textbf{0.1182} \\
\midrule
\multirow{4}{*}{SDS}
 & 0.05 & \textbf{19.48} & \textbf{0.2646} & \textbf{21.10} & 0.0800 \\
 & 0.10 & 19.14 & 0.2777 & 22.23 & 0.0608 \\
 & 0.15 & 18.74 & 0.2921 & 23.78 & 0.0642 \\
 & 0.20 & 18.49 & 0.3068 & 24.55 & \textbf{0.0964} \\
\bottomrule
\end{tabular}
\end{table}

\begin{table}[t]
\centering
\footnotesize
\setlength{\tabcolsep}{4pt}
\renewcommand{\arraystretch}{0.8}
\caption{Ablation study on perturbation budget $\epsilon$.
We report downstream editing results on SD~v1.5 Inpainting under a smaller $\ell_\infty$ budget ($\epsilon = 8/255$), using PSNR$\uparrow$, LPIPS$\downarrow$, and FID$\downarrow$. Best results within each protection are highlighted in bold.}
\label{tab:epsilon_ablation}
\begin{tabular}{llccc}
\toprule
\textbf{Protection} & \textbf{Purifier} & \textbf{PSNR$\uparrow$} & \textbf{LPIPS$\downarrow$} & \textbf{FID$\downarrow$} \\
\midrule
\multirow{6}{*}{AdvDM}
 & Unpurified    & 17.29 & 0.384 & 37.23 \\
 & JPEG        & 17.99 & 0.347 & 31.31 \\
 & IMPRESS     & 16.48 & 0.419 & 47.44 \\
 & GridPure    & 16.77 & 0.407 & 35.83 \\
 & \textbf{VAE-Trans} (ours)   & 17.22 & 0.379 & 27.46 \\
 & \textbf{EditorClean} (ours) & \textbf{19.39} & \textbf{0.266} & \textbf{20.08} \\
\midrule
\multirow{6}{*}{AdvPaint}
 & Unpurified    & 15.71 & 0.445 & 40.52 \\
 & JPEG        & 19.57 & 0.267 & 21.88 \\
 & IMPRESS     & 15.62 & 0.456 & 47.54 \\
 & GridPure    & 16.59 & 0.419 & 37.03 \\
 & \textbf{VAE-Trans} (ours)   & 17.65 & 0.359 & 26.24 \\
 & \textbf{EditorClean} (ours) & \textbf{19.68} & \textbf{0.255} & \textbf{19.33} \\
\midrule
\multirow{6}{*}{DiffusionGuard}
 & Unpurified    & 16.38 & 0.449 & 35.52 \\
 & JPEG        & \textbf{20.31} & \textbf{0.240} & \textbf{18.84} \\
 & IMPRESS     & 17.05 & 0.398 & 32.71 \\
 & GridPure    & 16.60 & 0.418 & 38.06 \\
 & \textbf{VAE-Trans} (ours)   & 17.73 & 0.354 & 25.92 \\
 & \textbf{EditorClean} (ours) & 19.43 & 0.264 & 19.99 \\
\midrule
\multirow{6}{*}{MIST}
 & Unpurified    & 16.13 & 0.449 & 43.00 \\
 & JPEG        & 18.40 & 0.327 & 28.49 \\
 & IMPRESS     & 15.48 & 0.488 & 58.06 \\
 & GridPure    & 16.44 & 0.426 & 37.38 \\
 & \textbf{VAE-Trans} (ours)   & 17.79 & 0.347 & 26.52 \\
 & \textbf{EditorClean} (ours) & \textbf{19.39} & \textbf{0.266} & \textbf{20.16} \\
\midrule
\multirow{6}{*}{PhotoGuard}
 & Unpurified    & 16.07 & 0.450 & 42.96 \\
 & JPEG        & 18.41 & 0.325 & 28.47 \\
 & IMPRESS     & 15.50 & 0.486 & 56.88 \\
 & GridPure    & 16.41 & 0.427 & 37.64 \\
 & \textbf{VAE-Trans} (ours)   & 17.79 & 0.346 & 26.19 \\
 & \textbf{EditorClean} (ours) & \textbf{19.38} & \textbf{0.267} & \textbf{20.45} \\
\midrule
\multirow{6}{*}{SDS}
 & Unpurified    & 17.33 & 0.410 & 32.54 \\
 & JPEG        & 19.08 & 0.300 & 24.63 \\
 & IMPRESS     & 16.78 & 0.434 & 36.57 \\
 & GridPure    & 16.50 & 0.421 & 36.16 \\
 & \textbf{VAE-Trans} (ours)   & 17.72 & 0.351 & 25.20 \\
 & \textbf{EditorClean} (ours) & \textbf{19.42} & \textbf{0.265} & \textbf{20.42} \\
\bottomrule
\end{tabular}
\end{table}

\subsection{Ablation Studies}\label{sec:ablation}

\noindent\textbf{Gaussian Noise Strength.}
We study the effect of Gaussian noise strength when combined with \textbf{EditorClean} purification. For each protection method, we add Gaussian noise with standard deviation $\sigma \in \{0.05, 0.10, 0.15, 0.20\}$ to the protected image before applying \textbf{EditorClean}. Quantitative results are reported in Table~\ref{tab:gaussian_ablation}, with qualitative examples shown in Figure~\ref{fig:gaussian_ablation_visual}.
The optimal noise level depends on the severity of the perturbations introduced by each protection method. Methods that impose relatively mild distortions, such as DiffusionGuard, AdvPaint, SDS, and AdvDM, achieve the best performance with lower noise levels ($\sigma = 0.05$ or $0.10$), while higher noise causes only minor degradation. In contrast, methods that introduce substantially stronger visual distortions, such as PhotoGuard and MIST, benefit from more aggressive noise injection: $\sigma = 0.15$ or $0.20$ yields both better quantitative results and visibly cleaner reconstructions.
Overall, these results indicate that the Gaussian noise strength should be calibrated to the severity of the visual distortions introduced by each protection method.
Protections that induce pronounced artifacts in the protected images benefit from stronger noise injection, whereas those with relatively mild distortions are better paired with lower noise levels.
Empirically, we find that $\sigma \approx 0.05$--$0.10$ works well for mild protections, while visually more aggressive protections favor $\sigma \approx 0.15$--$0.20$.
This trend is consistent with prior robustness studies~\cite{honig2024adversarial,zhao2024can} and is supported by our ablation results, where noise levels in this range consistently achieve strong purification performance without excessive loss of visual fidelity.

\noindent\textbf{Perturbation Budget.}
Our main experiments follow prior work and evaluate protection methods under a fixed $\ell_\infty$ budget of $\epsilon = 16/255$.
To examine sensitivity to perturbation strength, we repeat the downstream editing evaluation on SD~v1.5 Inpainting with a smaller budget, $\epsilon = 8/255$.
Table~\ref{tab:epsilon_ablation} reports PSNR$\uparrow$, LPIPS$\downarrow$, and FID$\downarrow$, and Figure~\ref{fig:epsilon_ablation_visual} in the appendix provides qualitative comparisons.
As expected, reducing the perturbation budget weakens the protections and improves editability across all settings.
However, the overall conclusions remain unchanged.
Purification, and \textbf{EditorClean} in particular, continues to largely neutralize the protection mechanisms.

\begin{table}[t]
\centering
\footnotesize
\setlength{\tabcolsep}{2pt}
\renewcommand{\arraystretch}{0.9}
\caption{Downstream editing results on Step1X-Edit (DiT). Protections are optimized against Step1X-Edit as the surrogate editor.
We report PSNR$\uparrow$, LPIPS$\downarrow$, and FID$\downarrow$ between Step1X-Edits from protected inputs and Step1X-Edits from clean inputs, and CLIP score$\uparrow$ for prompt-image alignment.
Best results within each protection are highlighted in bold.}
\label{tab:step1x_results}
\begin{tabular}{llcccc}
\toprule
\textbf{Protection} & \textbf{Purifier} & \textbf{PSNR$\uparrow$} & \textbf{LPIPS$\downarrow$} & \textbf{FID$\downarrow$} & \textbf{CLIP Score$\uparrow$} \\
\midrule
\multirow{4}{*}{PhotoGuard}
 & Unpurified         & 20.13 & 0.3360 & 117.87 & 21.28 \\
 & JPEG               & 16.85 & 0.3473 & 84.60 & 20.70 \\
 & GridPure           & 20.22 & 0.2841 & 84.24 & \textbf{21.33} \\
 & \textbf{EditorClean} (ours)  & \textbf{22.57} & \textbf{0.1992} & \textbf{56.85} & 20.31 \\
\midrule
\multirow{4}{*}{AdvDM}
 & Unpurified         & 17.81 & 0.4215 & 125.50 & 20.58 \\
 & JPEG               & 18.00 & 0.3174 & 78.55 & 20.77 \\
 & GridPure           & 20.58 & 0.2647 & 75.80 & \textbf{21.36} \\
 & \textbf{EditorClean} (ours)  & \textbf{22.81} & \textbf{0.1881} & \textbf{53.11} & 20.37 \\
\midrule
\multirow{4}{*}{MIST}
 & Unpurified         & 17.74 & 0.4192 & 124.46 & 20.80 \\
 & JPEG               & 17.90 & 0.3140 & 77.63 & 20.67 \\
 & GridPure           & 20.47 & 0.2660 & 74.88 & \textbf{21.35} \\
 & \textbf{EditorClean} (ours)  & \textbf{22.95} & \textbf{0.1749} & \textbf{51.73} & 20.49 \\
\midrule
Clean & -- & -- & -- & -- & 20.35 \\
\bottomrule
\end{tabular}
\end{table}

\subsection{Purification Under DiT Model Mismatch}
\label{sec:step1x}

This experiment investigates whether heterogeneity-driven purification persists when the downstream editor is DiT-based, ensuring the effect is not merely a byproduct of a UNet-to-DiT capability shift. We evaluate downstream editing on \textbf{Step1X-Edit}~\cite{liu2025step1x}, using prompt-only editing without inpainting masks to compare purified inputs against the clean-input Step1X baseline. We report PSNR$\uparrow$, LPIPS$\downarrow$, and FID$\downarrow$ for fidelity, alongside CLIP Score (CS)~\cite{radford2021learning}. CS measures the cosine similarity between text and image embeddings to reflect prompt-image alignment, though it does not directly measure edit fidelity or input consistency.
To ensure a fair comparison, we adapt \textbf{PhotoGuard}, \textbf{AdvDM}, and \textbf{MIST} to use a DiT editor as the optimization surrogate, generating perturbations directly against Step1X-Edit. These methods were selected because their optimization objectives are independent of UNet-specific components.

Table~\ref{tab:step1x_results} demonstrates that these protections remain effective on DiT-based editors; without purification, edits from protected inputs exhibit severe distortion. \textbf{EditorClean} achieves the most robust purification across all protections, reaching PSNR 22.57--22.95 and significantly reducing FID to 51.73--56.85. While \textbf{GridPure} attains the highest CLIP scores and improves all metrics over unpurified inputs, it consistently underperforms compared to \textbf{EditorClean} on similarity-based metrics.
Appendix Figure~\ref{fig:step1x_results_visual} provides representative qualitative examples corresponding to Table~\ref{tab:step1x_results}.
Finally, the performance of GridPure, which relies on a UNet denoising prior~\cite{zhao2024can}, against the DiT-based Step1X-Edit confirms that architectural heterogeneity is a powerful driver for purification. Furthermore, our \textbf{EditorClean} purifier is itself DiT-based (FLUX.1-fill-dev~\cite{labs2025flux}), proving that purification remains effective even in DiT-to-DiT scenarios as long as a model mismatch exists.

\begin{table}[t]
\centering
\footnotesize
\setlength{\tabcolsep}{10pt}
\renewcommand{\arraystretch}{0.9}
\caption{Impact of purification on DreamBooth fine-tuning. Lower FID and higher Precision indicate better generation quality.}
\label{tab:dreambooth_results}
\begin{tabular}{llcc}
\toprule
\textbf{Protection} & \textbf{Purifier} & \textbf{Precision$\uparrow$} & \textbf{FID$\downarrow$} \\
\midrule
\multirow{4}{*}{MIST}
 & Unpurified   & 0.347 & 176.08 \\
 & JPEG       & 0.479 & 148.76 \\
 & GridPure   & 0.542 & 143.80 \\
 & \textbf{EditorClean} & \textbf{0.613} & \textbf{137.85} \\
\midrule
\multirow{4}{*}{PhotoGuard}
 & Unpurified   & 0.237 & 200.94 \\
 & JPEG       & 0.429 & 150.44 \\
 & GridPure   & 0.547 & 147.01 \\
 & \textbf{EditorClean} & \textbf{0.613} & \textbf{141.82} \\
\midrule
\multirow{4}{*}{Unprotected}
 & JPEG       & 0.584 & 136.91 \\
 & GridPure   & 0.618 & 140.46 \\
 & \textbf{EditorClean} & 0.613 & 132.09 \\
 & --         & 0.592 & 135.66 \\
\bottomrule
\end{tabular}
\end{table}

\begin{table}[t]
\centering
\footnotesize
\setlength{\tabcolsep}{8pt}
\renewcommand{\arraystretch}{0.9}
\caption{Impact of purification on textual inversion style mimicry ($\epsilon = 8/255$). We report CLIP Acc$\uparrow$ and FID$\downarrow$.}
\label{tab:textual_inversion_results}

\begin{tabular}{llcc}
\toprule
\textbf{Protection} & \textbf{Purifier} & \textbf{CLIP Acc$\uparrow$} & \textbf{FID$\downarrow$} \\
\midrule
\multirow{4}{*}{AdvDM}
 & Unpurified        & 0.090 & 517.78 \\
 & JPEG            & 0.660 & 407.70 \\
 & GridPure        & 0.937 & 423.60 \\
 & \textbf{EditorClean} (ours) & \textbf{0.997} & \textbf{317.79} \\
\midrule
\multirow{4}{*}{MIST}
 & Unpurified        & 0.007 & 470.41 \\
 & JPEG            & 0.903 & 360.79 \\
 & GridPure        & 0.910 & 389.70 \\
 & \textbf{EditorClean} (ours) & \textbf{0.993} & \textbf{315.78} \\
\midrule
\multirow{4}{*}{PhotoGuard}
 & Unpurified        & 0.123 & 454.97 \\
 & JPEG            & 0.923 & 358.64 \\
 & GridPure        & 0.883 & 363.49 \\
 & \textbf{EditorClean} (ours) & \textbf{0.977} & \textbf{320.17} \\
\midrule
\multirow{4}{*}{SDS}
 & Unpurified        & 0.167 & 440.21 \\
 & JPEG            & 0.647 & 429.29 \\
 & GridPure        & 0.767 & 426.02 \\
 & \textbf{EditorClean} (ours) & \textbf{0.997} & \textbf{302.31} \\
\midrule
\multirow{4}{*}{Unprotected}
 & JPEG            & 1.000 & 315.31 \\
 & GridPure        & 0.943 & 392.75 \\
 & \textbf{EditorClean} (ours) & 0.987 & 306.92 \\
 & --              & 0.997 & 300.51 \\
\bottomrule
\end{tabular}
\end{table}

\subsection{Purification of Unlearnable Samples}
\label{sec:unlearnable}

We evaluate whether purification can restore subject-driven fine-tuning on unlearnable samples.
This setting is particularly relevant to artistic style protection, which aims to prevent model personalization.
Among six protection methods, we focus on PhotoGuard and MIST.
Both are designed to disrupt DreamBooth-based style personalization.
PhotoGuard introduces latent distortions that hinder fine-tuning convergence.
MIST uses texture-aware objectives to prevent style imitation during adaptation.
We conduct two evaluations to test whether purification can recover subject-driven fine-tuning and style mimicry on protected images.

\noindent\textbf{DreamBooth Personalization.}
We fine-tune Stable Diffusion v1.5 on 19 subject categories from the DreamBooth dataset~\cite{ruiz2023dreambooth} using the official Diffusers script.
We report FID~\cite{heusel2017gans} and Precision~\cite{kynkaanniemi2019improved}, where Precision is the fraction of generated images classified as the target category.
Hyperparameters are given in Appendix~\ref{sec:unlearnable_hyperparams}.

\noindent\textbf{Textual Inversion Style Mimicry.}
We further evaluate style protection using textual inversion on artworks from three artists~\cite{gal2022image}.
Following prior work~\cite{cao2023impress}, we measure style mimicry with CLIP-based style classification accuracy, where a pretrained CLIP model~\cite{radford2021learning} predicts art genres from generated images.

Table~\ref{tab:dreambooth_results} shows that protection methods severely degrade DreamBooth performance, while \textbf{EditorClean} largely restores generation quality to near-clean levels.
Table~\ref{tab:textual_inversion_results} shows a similar trend.
We report CLIP-based style classification accuracy (CLIP Acc$\uparrow$) and generation quality (FID$\downarrow$), averaged over three artists with 100 generated images each.
Protections sharply reduce CLIP Acc and increase FID, but \textbf{EditorClean} recovers both metrics.
In contrast, JPEG and GridPure provide only limited gains.
Overall, these results show that \textbf{EditorClean} generalizes across tasks and effectively neutralizes protections against subject-driven personalization and artistic style mimicry.

\section{Discussion}

\begin{figure}[t]
\centering
\includegraphics[width=\columnwidth]{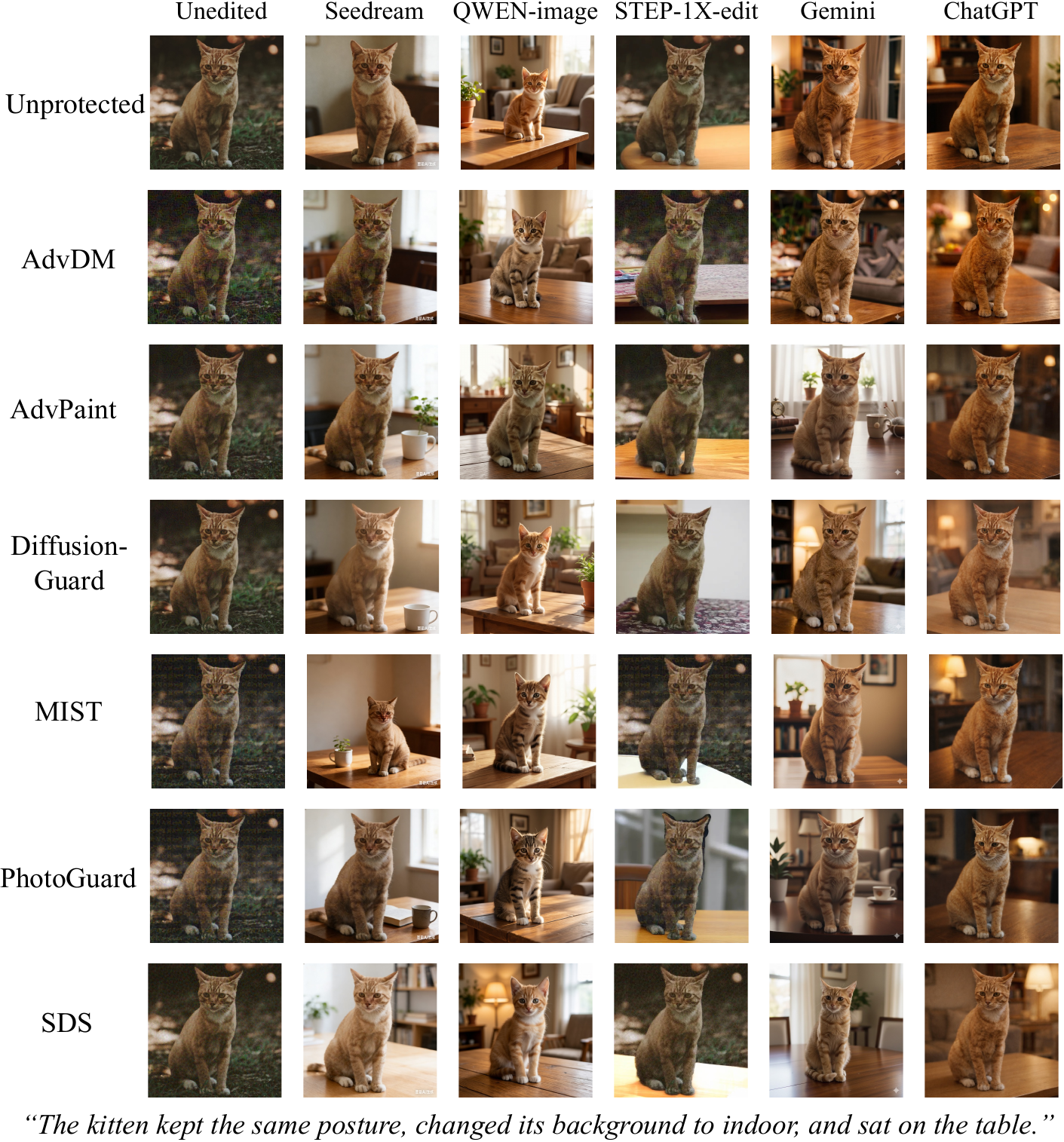}
\caption{Real-world editor comparison under model mismatch.}
\label{fig:cross_arch_llm}
\end{figure}

\noindent\textbf{Real-World Editors under Model Mismatch.}

Beyond our main benchmark, we further compare how protections behave on a diverse set of real-world editing platforms that users may adopt in practice under the same SD-surrogate threat model.
All protected inputs in this experiment are adversarial examples generated by optimizing perturbations against SD~v1.5 Inpainting, and then directly evaluated on heterogeneous editors.
These platforms include \textbf{Seedream (Doubao)}~\cite{seedream2025seedream}, \textbf{Qwen-Image}~\cite{wu2025qwen}, \textbf{Step1X-Edit}~\cite{liu2025step1x}, \textbf{ChatGPT-4o}~\cite{cao2025preliminary}, and \textbf{Gemini Pro}~\cite{zuo2025nano}.
They represent a diverse ecosystem of widely used editing systems.
All platforms successfully perform semantic edits on protected images with quality comparable to clean inputs.
This result confirms our central claim: current protection methods remain effective only in matched-surrogate settings where the target editor matches the optimization surrogate.

Figure~\ref{fig:cross_arch_llm} shows representative results using the same editing instruction as in our SD-based benchmark, while extended results with a more complex cartoon example are provided in Appendix Figure~\ref{fig:cross_arch_llm_appendix}.
These results highlight that the vulnerability arises in realistic downstream usage, where the attacker can simply adopt a different editor from the open ecosystem.
\textbf{EditorClean} provides a practical and accessible attack vector that does not depend on commercial API access.
As shown in Table~\ref{tab:edited_result_comparison}, the purified outputs approach clean image baselines.
They achieve PSNR above 18 dB and FID below 25 on most defenses.
These results confirm that even compact and publicly trainable purifiers can largely neutralize existing protection methods.

\noindent\textbf{Implications for Defenses and Evaluation.}
Our results suggest a largely \emph{purify-once, edit-freely} failure mode: once a protected image has been successfully purified, either explicitly via a purification operator $\mathcal{P}$ or implicitly via a mismatched editor's reconstruction prior, the protective perturbation is largely erased and subsequent edits can proceed without restriction.

These findings motivate protection mechanisms that remain robust under heterogeneous attackers, rather than relying on a single surrogate editor or a fixed reconstruction prior.
At a minimum, evaluations should include model-family and architecture mismatch, as well as realistic preprocessing and reconstruction pipelines after release.
More broadly, perturbation-based protections should be treated as one layer in a defense-in-depth strategy, complemented by provenance and traceability mechanisms and platform-side enforcement.
We also note that diffusion-based purification can neutralize adversarial perturbations beyond generative editing; our extension to ImageNet classification under AutoAttack further supports this perspective (Appendix~\ref{sec:autoattack}).

\section{Conclusion}

We present a unified post-release purification framework to evaluate the survivability of perturbation-based image protections under model mismatch, and introduce two practical purifiers: \textbf{VAE-Trans}, based on latent-space projection, and \textbf{EditorClean}, which performs instruction-guided reconstruction using an alternative editing model.
Our evaluation reveals a critical \emph{purify-once, edit-freely} failure mode: once protective signals are removed by a mismatched pipeline, images remain unprotected for subsequent edits.
Notably, \textbf{EditorClean} consistently restores editability, achieving 3--6\,dB PSNR gains and 50--70\% FID reductions across six defenses.
These findings show that existing protections are highly vulnerable to heterogeneous post-release attackers, underscoring the need for defenses with genuine cross-model robustness.

\cleardoublepage
\section*{Ethical Considerations}

\noindent\textbf{Scope and dual use.}
Our study evaluates how post-release purification under model mismatch can suppress or remove perturbation-based image protections.
While intended to improve evaluation practice, this capability can be misused to bypass creator-intended safeguards and enable unauthorized editing or model training, including nonconsensual imagery, identity manipulation, disinformation, and unauthorized style imitation.

\noindent\textbf{Stakeholders.}
The primary stakeholders include (i)~content creators and rights holders who deploy proactive protections, (ii)~individuals depicted in images (including public figures) whose identity may be manipulated, (iii)~downstream platforms and editing services that may be used to scale misuse, and (iv)~researchers and practitioners who rely on accurate robustness claims when deploying protections.

\noindent\textbf{Methodological mitigations (implemented).}
We conduct experiments using only publicly available models and datasets.
The DiffusionGuard benchmark contains publicly sourced images and is used solely for robustness evaluation; we report results in aggregate and do not release any private data.
For commercial editors, we submit only benchmark images through standard user-facing interfaces, do not attempt to bypass safety mechanisms, and do not probe or extract proprietary details.
To reduce misuse risk while enabling reproducibility, our released artifact focuses on reproducing the evaluation pipeline and does not include third-party model weights or commercial API automation; users must obtain external weights separately and comply with their licenses.

\noindent\textbf{Recommended deployment measures.}
Our findings underscore that perturbation-based protections should not be treated as a standalone safeguard.
Practical deployments should pair them with provenance and traceability mechanisms, platform-side policy enforcement and monitoring, and human oversight in high-stakes contexts.

\noindent\textbf{Justification.}
Characterizing failure modes under realistic post-release pipelines helps avoid a false sense of security and informs more robust defense design and evaluation protocols.
We therefore view responsible disclosure of these results as necessary to support safer deployment of diffusion-based editing systems.

\cleardoublepage

\section*{Open Science}
To facilitate reproducibility, we will publicly release the full artifact package accompanying this paper.
The release will include our proposed purifiers (\textbf{VAE-Trans} and \textbf{EditorClean}), integrated baselines (JPEG/IMPRESS/GridPure), and evaluation scripts for generating protected inputs and assessing downstream editability under the matched-surrogate, cross-version, and purification-then-edit settings described in this work.
The package will also provide the DiffusionGuard benchmark images, masks, and prompts used in Section~\ref{sec:experiments}, along with scripts for downloading and preparing the OmniEdit-Filtered-1.2M dataset used for purifier training.

\noindent\textbf{Reproducibility.}
The artifact is organized into submodules for (i)~the main SD-based benchmark (protection generation and downstream inpainting evaluation), (ii)~purification baselines, (iii)~\textbf{EditorClean}, (iv)~\textbf{VAE-Trans}, and (v)~the DiT-based Step1X evaluation.
The repository documentation provides step-by-step instructions and a point-by-point mapping from each table/figure in the paper to the corresponding scripts and configurations needed to reproduce it.

\noindent\textbf{Environment and dependencies.}
We provide a pip-based environment setup (Python~3.10+; a CUDA-capable GPU is recommended).
For lower VRAM, we support optional GGUF quantized weights for FLUX.1-fill-dev.
Some baselines rely on separate environments as documented in the artifact.

\noindent\textbf{EditorClean hyperparameters.}
We include the full training configuration used in our experiments; key settings are:
backbone \texttt{black-forest-labs/FLUX.1-Fill-dev} (dtype \texttt{bfloat16});
denoising training data drawn from OmniEdit-Filtered-1.2M with 512$\times$512 images, \texttt{drop\_text\_prob}=0.1, and synthetic Gaussian noise with \texttt{noise\_level}=0.1;
LoRA rank $r=32$ and $\alpha=32$ (Gaussian init) applied to the Flux embedder, attention projections (Q/K/V/out), and MLP output layers;
optimizer Prodigy (\texttt{lr}=1, \texttt{weight\_decay}=0.01);
batch size 2, gradient checkpointing enabled, and 2,000 training steps with a fixed seed (666).

\noindent\textbf{External dependencies and variability.}
Reproducing some experiments requires separately downloading third-party model weights (e.g., Stable Diffusion, FLUX.1-fill-dev, and Step1X-Edit) and complying with their licenses.
Results involving commercial editing services may vary over time.

\cleardoublepage
\bibliographystyle{plainurl}
\bibliography{ref}

@String{Computer = "{IEEE} Computer" }

@inproceedings{rombach2022high,
  title={High-resolution image synthesis with latent diffusion models},
  author={Rombach, Robin and Blattmann, Andreas and Lorenz, Dominik and Esser, Patrick and Ommer, Bj{\"o}rn},
  booktitle={Proceedings of the IEEE/CVF Conference on Computer Vision and Pattern Recognition},
  pages={10684--10695},
  year={2022}
}

@article{ramesh2022hierarchical,
  title={Hierarchical text-conditional image generation with clip latents},
  author={Ramesh, Aditya and Dhariwal, Prafulla and Nichol, Alex and Chu, Casey and Chen, Mark},
  journal={arXiv preprint arXiv:2204.06125},
  year={2022}
}

@article{saharia2022photorealistic,
  title={Photorealistic text-to-image diffusion models with deep language understanding},
  author={Saharia, Chitwan and Chan, William and Saxena, Saurabh and Li, Lala and Whang, Jay and Denton, Emily L and others},
  journal={Advances in Neural Information Processing Systems},
  volume={35},
  pages={36479--36494},
  year={2022}
}

@article{hertz2022prompt,
  title={Prompt-to-prompt image editing with cross attention control},
  author={Hertz, Amir and Mokady, Ron and Tenenbaum, Jay and Aberman, Kfir and Pritch, Yael and Cohen-Or, Daniel},
  journal={arXiv preprint arXiv:2208.01626},
  year={2022}
}

@inproceedings{zhang2023adding,
  title={Adding conditional control to text-to-image diffusion models},
  author={Zhang, Lvmin and Rao, Anyi and Agrawala, Maneesh},
  booktitle={Proceedings of the IEEE/CVF international conference on computer vision},
  pages={3836--3847},
  year={2023}
}

@inproceedings{brooks2023instructpix2pix,
  title={Instructpix2pix: Learning to follow image editing instructions},
  author={Brooks, Tim and Holynski, Aleksander and Efros, Alexei A},
  booktitle={Proceedings of the IEEE/CVF conference on computer vision and pattern recognition},
  pages={18392--18402},
  year={2023}
}

@inproceedings{salman2023raising,
  title={Raising the Cost of Malicious AI-Powered Image Editing},
  author={Salman, Hadi and Khaddaj, Alaa and Leclerc, Guillaume and Ilyas, Andrew and Madry, Aleksander},
  booktitle={International Conference on Machine Learning},
  pages={29894--29918},
  year={2023},
  organization={PMLR}
}

@inproceedings{shan2023glaze,
  title={Glaze: Protecting artists from style mimicry by $\{$Text-to-Image$\}$ models},
  author={Shan, Shawn and Cryan, Jenna and Wenger, Emily and Zheng, Haitao and Hanocka, Rana and Zhao, Ben Y},
  booktitle={32nd USENIX Security Symposium (USENIX Security 23)},
  pages={2187--2204},
  year={2023}
}

@article{liang2023mist,
  title={Mist: Towards improved adversarial examples for diffusion models},
  author={Liang, Chumeng and Wu, Xiaoyu},
  journal={arXiv preprint arXiv:2305.12683},
  year={2023}
}

@inproceedings{lo2024distraction,
  title={Distraction is all you need: Memory-efficient image immunization against diffusion-based image editing},
  author={Lo, Ling and Yeo, Cheng Yu and Shuai, Hong-Han and Cheng, Wen-Huang},
  booktitle={Proceedings of the IEEE/CVF Conference on Computer Vision and Pattern Recognition},
  pages={24462--24471},
  year={2024}
}

@article{honig2024adversarial,
  title={Adversarial perturbations cannot reliably protect artists from generative ai},
  author={H{\"o}nig, Robert and Rando, Javier and Carlini, Nicholas and Tram{\`e}r, Florian},
  journal={arXiv preprint arXiv:2406.12027},
  year={2024}
}

@article{cao2023impress,
  title={Impress: Evaluating the resilience of imperceptible perturbations against unauthorized data usage in diffusion-based generative ai},
  author={Cao, Bochuan and Li, Changjiang and Wang, Ting and Jia, Jinyuan and Li, Bo and Chen, Jinghui},
  journal={Advances in Neural Information Processing Systems},
  volume={36},
  pages={10657--10677},
  year={2023}
}

@article{madry2017towards,
  title={Towards deep learning models resistant to adversarial attacks},
  author={Madry, Aleksander and Makelov, Aleksandar and Schmidt, Ludwig and Tsipras, Dimitris and Vladu, Adrian},
  journal={arXiv preprint arXiv:1706.06083},
  year={2017}
}

@inproceedings{nie2022diffusion,
  title={Diffusion Models for Adversarial Purification},
  author={Nie, Weili and Guo, Brandon and Huang, Yujia and Xiao, Chaowei and Vahdat, Arash and Anandkumar, Animashree},
  booktitle={International Conference on Machine Learning},
  pages={16805--16827},
  year={2022},
  organization={PMLR}
}

@inproceedings{zhang2018unreasonable,
  title={The unreasonable effectiveness of deep features as a perceptual metric},
  author={Zhang, Richard and Isola, Phillip and Efros, Alexei A and Shechtman, Eli and Wang, Oliver},
  booktitle={Proceedings of the IEEE conference on computer vision and pattern recognition},
  pages={586--595},
  year={2018}
}

@article{meng2021sdedit,
  title={Sdedit: Guided image synthesis and editing with stochastic differential equations},
  author={Meng, Chenlin and He, Yutong and Song, Yang and Song, Jiaming and Wu, Jiajun and Zhu, Jun-Yan and Ermon, Stefano},
  journal={arXiv preprint arXiv:2108.01073},
  year={2021}
}

@inproceedings{radford2021learning,
  title={Learning transferable visual models from natural language supervision},
  author={Radford, Alec and Kim, Jong Wook and Hallacy, Chris and Ramesh, Aditya and Goh, Gabriel and Agarwal, Sandhini and Sastry, Girish and Askell, Amanda and Mishkin, Pamela and Clark, Jack and others},
  booktitle={International conference on machine learning},
  pages={8748--8763},
  year={2021},
  organization={PmLR}
}

@article{ho2020denoising,
  title={Denoising diffusion probabilistic models},
  author={Ho, Jonathan and Jain, Ajay and Abbeel, Pieter},
  journal={Advances in neural information processing systems},
  volume={33},
  pages={6840--6851},
  year={2020}
}

@inproceedings{sohl2015deep,
  title={Deep unsupervised learning using nonequilibrium thermodynamics},
  author={Sohl-Dickstein, Jascha and Weiss, Eric and Maheswaranathan, Niru and Ganguli, Surya},
  booktitle={International conference on machine learning},
  pages={2256--2265},
  year={2015},
  organization={pmlr}
}

@inproceedings{DBLP:conf/iclr/XueLWC24,
  author       = {Haotian Xue and
                  Chumeng Liang and
                  Xiaoyu Wu and
                  Yongxin Chen},
  title        = {Toward effective protection against diffusion-based mimicry through
                  score distillation},
  booktitle    = {The Twelfth International Conference on Learning Representations,
                  {ICLR} 2024, Vienna, Austria, May 7-11, 2024},
  publisher    = {OpenReview.net},
  year         = {2024},
  url          = {https://openreview.net/forum?id=NzxCMe88HX},
  bibsource    = {dblp computer science bibliography, https://dblp.org}
}

@inproceedings{liang2023adversarial,
  title={Adversarial example does good: preventing painting imitation from diffusion models via adversarial examples},
  author={Liang, Chumeng and Wu, Xiaoyu and Hua, Yang and Zhang, Jiaru and Xue, Yiming and Song, Tao and Xue, Zhengui and Ma, Ruhui and Guan, Haibing},
  booktitle={40th International Conference on Machine Learning, ICML 2023},
  pages={20763--20786},
  year={2023}
}

@article{choi2024diffusionguard,
  title={Diffusionguard: A robust defense against malicious diffusion-based image editing},
  author={Choi, June Suk and Lee, Kyungmin and Jeong, Jongheon and Xie, Saining and Shin, Jinwoo and Lee, Kimin},
  journal={arXiv preprint arXiv:2410.05694},
  year={2024}
}

@misc{jeon2025advpaintprotectingimagesinpainting,
      title={AdvPaint: Protecting Images from Inpainting Manipulation via Adversarial Attention Disruption},
      author={Joonsung Jeon and Woo Jae Kim and Suhyeon Ha and Sooel Son and Sung-eui Yoon},
      year={2025},
      eprint={2503.10081},
      archivePrefix={arXiv},
      primaryClass={cs.CV},
      url={https://arxiv.org/abs/2503.10081},
}

@inproceedings{zhao2024can,
  title={Can protective perturbation safeguard personal data from being exploited by stable diffusion?},
  author={Zhao, Zhengyue and Duan, Jinhao and Xu, Kaidi and Wang, Chenan and Zhang, Rui and Du, Zidong and Guo, Qi and Hu, Xing},
  booktitle={Proceedings of the IEEE/CVF Conference on Computer Vision and Pattern Recognition},
  pages={24398--24407},
  year={2024}
}

@article{heusel2017gans,
  title={Gans trained by a two time-scale update rule converge to a local nash equilibrium},
  author={Heusel, Martin and Ramsauer, Hubert and Unterthiner, Thomas and Nessler, Bernhard and Hochreiter, Sepp},
  journal={Advances in neural information processing systems},
  volume={30},
  year={2017}
}

@article{xu2023imagereward,
  title={Imagereward: Learning and evaluating human preferences for text-to-image generation},
  author={Xu, Jiazheng and Liu, Xiao and Wu, Yuchen and Tong, Yuxuan and Li, Qinkai and Ding, Ming and Tang, Jie and Dong, Yuxiao},
  journal={Advances in Neural Information Processing Systems},
  volume={36},
  pages={15903--15935},
  year={2023}
}

@article{dhariwal2021diffusion,
  title={Diffusion models beat gans on image synthesis},
  author={Dhariwal, Prafulla and Nichol, Alexander},
  journal={Advances in neural information processing systems},
  volume={34},
  pages={8780--8794},
  year={2021}
}

@article{croitoru2023diffusion,
  title={Diffusion models in vision: A survey},
  author={Croitoru, Florinel-Alin and Hondru, Vlad and Ionescu, Radu Tudor and Shah, Mubarak},
  journal={IEEE transactions on pattern analysis and machine intelligence},
  volume={45},
  number={9},
  pages={10850--10869},
  year={2023},
  publisher={Ieee}
}

@inproceedings{ruiz2023dreambooth,
  title={Dreambooth: Fine tuning text-to-image diffusion models for subject-driven generation},
  author={Ruiz, Nataniel and Li, Yuanzhen and Jampani, Varun and Pritch, Yael and Rubinstein, Michael and Aberman, Kfir},
  booktitle={Proceedings of the IEEE/CVF conference on computer vision and pattern recognition},
  pages={22500--22510},
  year={2023}
}

@inproceedings{van2023anti,
  title={Anti-dreambooth: Protecting users from personalized text-to-image synthesis},
  author={Van Le, Thanh and Phung, Hao and Nguyen, Thuan Hoang and Dao, Quan and Tran, Ngoc N and Tran, Anh},
  booktitle={Proceedings of the IEEE/CVF International Conference on Computer Vision},
  pages={2116--2127},
  year={2023}
}

@inproceedings{peebles2023scalable,
  title={Scalable diffusion models with transformers},
  author={Peebles, William and Xie, Saining},
  booktitle={Proceedings of the IEEE/CVF international conference on computer vision},
  pages={4195--4205},
  year={2023}
}

@article{hu2022lora,
  title={Lora: Low-rank adaptation of large language models.},
  author={Hu, Edward J and Shen, Yelong and Wallis, Phillip and Allen-Zhu, Zeyuan and Li, Yuanzhi and Wang, Shean and Wang, Lu and Chen, Weizhu and others},
  journal={ICLR},
  volume={1},
  number={2},
  pages={3},
  year={2022}
}

@article{loshchilov2017decoupled,
  title={Decoupled weight decay regularization},
  author={Loshchilov, Ilya and Hutter, Frank},
  journal={arXiv preprint arXiv:1711.05101},
  year={2017}
}

@article{huang2021unlearnable,
  title={Unlearnable examples: Making personal data unexploitable},
  author={Huang, Hanxun and Ma, Xingjun and Erfani, Sarah Monazam and Bailey, James and Wang, Yisen},
  journal={arXiv preprint arXiv:2101.04898},
  year={2021}
}

@article{fu2022robust,
  title={Robust unlearnable examples: Protecting data against adversarial learning},
  author={Fu, Shaopeng and He, Fengxiang and Liu, Yang and Shen, Li and Tao, Dacheng},
  journal={arXiv preprint arXiv:2203.14533},
  year={2022}
}

@article{cherepanova2021lowkey,
  title={Lowkey: Leveraging adversarial attacks to protect social media users from facial recognition},
  author={Cherepanova, Valeriia and Goldblum, Micah and Foley, Harrison and Duan, Shiyuan and Dickerson, John and Taylor, Gavin and Goldstein, Tom},
  journal={arXiv preprint arXiv:2101.07922},
  year={2021}
}

@article{sandoval2023jpeg,
  title={Jpeg compressed images can bypass protections against ai editing},
  author={Sandoval-Segura, Pedro and Geiping, Jonas and Goldstein, Tom},
  journal={arXiv preprint arXiv:2304.02234},
  year={2023}
}

@article{wallace1991jpeg,
  title={The JPEG still picture compression standard},
  author={Wallace, Gregory K},
  journal={Communications of the ACM},
  volume={34},
  number={4},
  pages={30--44},
  year={1991},
  publisher={AcM New York, NY, USA}
}

@article{wu2025qwen,
  title={Qwen-image technical report},
  author={Wu, Chenfei and Li, Jiahao and Zhou, Jingren and Lin, Junyang and Gao, Kaiyuan and Yan, Kun and Yin, Sheng-ming and Bai, Shuai and Xu, Xiao and Chen, Yilei and others},
  journal={arXiv preprint arXiv:2508.02324},
  year={2025}
}

@article{liu2025step1x,
  title={Step1x-edit: A practical framework for general image editing},
  author={Liu, Shiyu and Han, Yucheng and Xing, Peng and Yin, Fukun and Wang, Rui and Cheng, Wei and Liao, Jiaqi and Wang, Yingming and Fu, Honghao and Han, Chunrui and others},
  journal={arXiv preprint arXiv:2504.17761},
  year={2025}
}

@article{seedream2025seedream,
  title={Seedream 4.0: Toward next-generation multimodal image generation},
  author={Seedream, Team and Chen, Yunpeng and Gao, Yu and Gong, Lixue and Guo, Meng and Guo, Qiushan and Guo, Zhiyao and Hou, Xiaoxia and Huang, Weilin and Huang, Yixuan and others},
  journal={arXiv preprint arXiv:2509.20427},
  year={2025}
}

@article{cao2025preliminary,
  title={Preliminary Explorations with GPT-4o (mni) Native Image Generation},
  author={Cao, Pu and Zhou, Feng and Ji, Junyi and Kong, Qingye and Lv, Zhixiang and Zhang, Mingjian and Zhao, Xuekun and Wu, Siqi and Lin, Yinghui and Song, Qing and others},
  journal={arXiv preprint arXiv:2505.05501},
  year={2025}
}

@article{zuo2025nano,
  title={Is Nano Banana Pro a Low-Level Vision All-Rounder? A Comprehensive Evaluation on 14 Tasks and 40 Datasets},
  author={Zuo, Jialong and Deng, Haoyou and Zhou, Hanyu and Zhu, Jiaxin and Zhang, Yicheng and Zhang, Yiwei and Yan, Yongxin and Huang, Kaixing and Chen, Weisen and Deng, Yongtai and others},
  journal={arXiv preprint arXiv:2512.15110},
  year={2025}
}

@inproceedings{zhang2025enabling,
  title={Enabling Instructional Image Editing with In-Context Generation in Large Scale Diffusion Transformer},
  author={Zhang, Zechuan and Xie, Ji and Lu, Yu and Yang, Zongxin and Yang, Yi},
  booktitle={The Thirty-ninth Annual Conference on Neural Information Processing Systems},
  year={2025}
}

@article{kynkaanniemi2019improved,
  title={Improved precision and recall metric for assessing generative models},
  author={Kynk{\"a}{\"a}nniemi, Tuomas and Karras, Tero and Laine, Samuli and Lehtinen, Jaakko and Aila, Timo},
  journal={Advances in neural information processing systems},
  volume={32},
  year={2019}
}

@article{labs2025flux,
  title={FLUX. 1 Kontext: Flow Matching for In-Context Image Generation and Editing in Latent Space},
  author={Labs, Black Forest and Batifol, Stephen and Blattmann, Andreas and Boesel, Frederic and Consul, Saksham and Diagne, Cyril and Dockhorn, Tim and English, Jack and English, Zion and Esser, Patrick and others},
  journal={arXiv preprint arXiv:2506.15742},
  year={2025}
}

@inproceedings{wei2024omniedit,
  title={Omniedit: Building image editing generalist models through specialist supervision},
  author={Wei, Cong and Xiong, Zheyang and Ren, Weiming and Du, Xeron and Zhang, Ge and Chen, Wenhu},
  booktitle={The Thirteenth International Conference on Learning Representations},
  year={2024}
}

@article{song2020denoising,
  title={Denoising diffusion implicit models},
  author={Song, Jiaming and Meng, Chenlin and Ermon, Stefano},
  journal={arXiv preprint arXiv:2010.02502},
  year={2020}
}

@article{ho2022classifier,
  title={Classifier-free diffusion guidance},
  author={Ho, Jonathan and Salimans, Tim},
  journal={arXiv preprint arXiv:2207.12598},
  year={2022}
}

@inproceedings{foerster2025lightshed,
  title={$\{$LightShed$\}$: Defeating Perturbation-based Image Copyright Protections},
  author={Foerster, Hanna and Behrouzi, Sasha and Rieger, Phillip and Jadliwala, Murtuza and Sadeghi, Ahmad-Reza},
  booktitle={34th USENIX Security Symposium (USENIX Security 25)},
  pages={7271--7290},
  year={2025}
}

@inproceedings{croce2020reliable,
  title={Reliable evaluation of adversarial robustness with an ensemble of diverse parameter-free attacks},
  author={Croce, Francesco and Hein, Matthias},
  booktitle={International conference on machine learning},
  pages={2206--2216},
  year={2020},
  organization={PMLR}
}

@inproceedings{gal2022image,
  title={An image is worth one word: Personalizing text-to-image generation using textual inversion},
  author={Gal, Rinon and Alaluf, Yuval and Atzmon, Yuval and Patashnik, Or and Bermano, Amit H and Chechik, Gal and Cohen-Or, Daniel},
  booktitle={International Conference on Learning Representations},
  year={2023}
}

\cleardoublepage
\appendix

\section{Extension to Classification Tasks}
\label{sec:autoattack}

Our primary focus is the evaluation of the survivability of perturbation-based protections designed for diffusion-based generative models under purification after release.
To examine whether our purification framework generalizes beyond the generative editing domain, we extend the evaluation to adversarial perturbations crafted for standard image classification tasks.
This analysis has two goals.
First, it shows that diffusion-based purification neutralizes adversarial perturbations across different threat models, not only those targeting generative pipelines.
Second, it verifies that the reconstruction capability of \textbf{EditorClean} transfers to classical adversarial robustness settings.

We evaluate our purification methods under the widely adopted AutoAttack benchmark~\cite{croce2020reliable}, which provides a standardized protocol for adversarial robustness evaluation in image classification.
We follow the evaluation protocol of DiffPure~\cite{nie2022diffusion} (Table~8) and consider a standard non-adaptive transfer setting.
Specifically, we generate $\ell_\infty$-bounded adversarial examples against a fixed ImageNet classifier, ResNet-50, using the AutoAttack standard suite, which includes APGD-CE (Auto-PGD with cross-entropy loss), APGD-DLR (Auto-PGD with difference of logits ratio loss), FAB (Fast Adaptive Boundary attack), and Square (a score-based black-box attack).
Following standard adversarial robustness evaluation protocols~\cite{croce2020reliable,nie2022diffusion}, we report two complementary metrics: \emph{standard accuracy}, which measures classification accuracy on clean images and quantifies any accuracy degradation introduced by the purification process, and \emph{robust accuracy}, which measures classification accuracy on adversarial examples and directly reflects defense effectiveness.
We evaluate each purifier by passing the adversarial images through a purifier and classifier pipeline.
We do not differentiate through the purifier during evaluation.
We adopt the same $\ell_\infty$ threat model as in the ImageNet setting of DiffPure~\cite{nie2022diffusion}, with a perturbation budget of $\epsilon = 4/255$, and evaluate on 1,024 randomly sampled ImageNet validation images.
For DiffPure, we use the default ImageNet diffusion timestep $t^{*}=0.15$, as reported in Table~3 of~\cite{nie2022diffusion}.
For \textbf{EditorClean}, we follow the same inference configuration as in Section~\ref{sec:editorclean}, including test-time noise injection with $\sigma_{\mathrm{test}}=0.10$.
As shown in Table~\ref{tab:autoattack_imagenet}, AutoAttack fully compromises the static classifier under this protocol and yields 0\% robust accuracy.
In contrast, both DiffPure and \textbf{EditorClean} substantially restore robustness.
\textbf{EditorClean} achieves the best trade-off, with 65.6\% robust accuracy and 69.0\% standard accuracy on the sampled subset.

\begin{table}[t]
\centering
\footnotesize
\setlength{\tabcolsep}{10pt}
\caption{Standard accuracy and robust accuracy under AutoAttack $\ell_\infty$ ( $\epsilon = 4/255$ ) on ImageNet (ResNet-50), evaluated on 1,024 randomly sampled validation images following the transfer-based protocol of DiffPure.}
\label{tab:autoattack_imagenet}
\begin{tabular}{lcc}
\toprule
\textbf{Purifier} & \textbf{Clean Acc (\%)} & \textbf{Robust Acc (\%)} \\
\midrule
Unpurified & 78.5 & 0.0 \\
DiffPure & 64.4 & 63.2 \\
\textbf{EditorClean} (ours) & 69.0 & 65.6 \\
\bottomrule
\end{tabular}
\end{table}

\begin{figure*}[t]
\centering
\includegraphics[width=\textwidth]{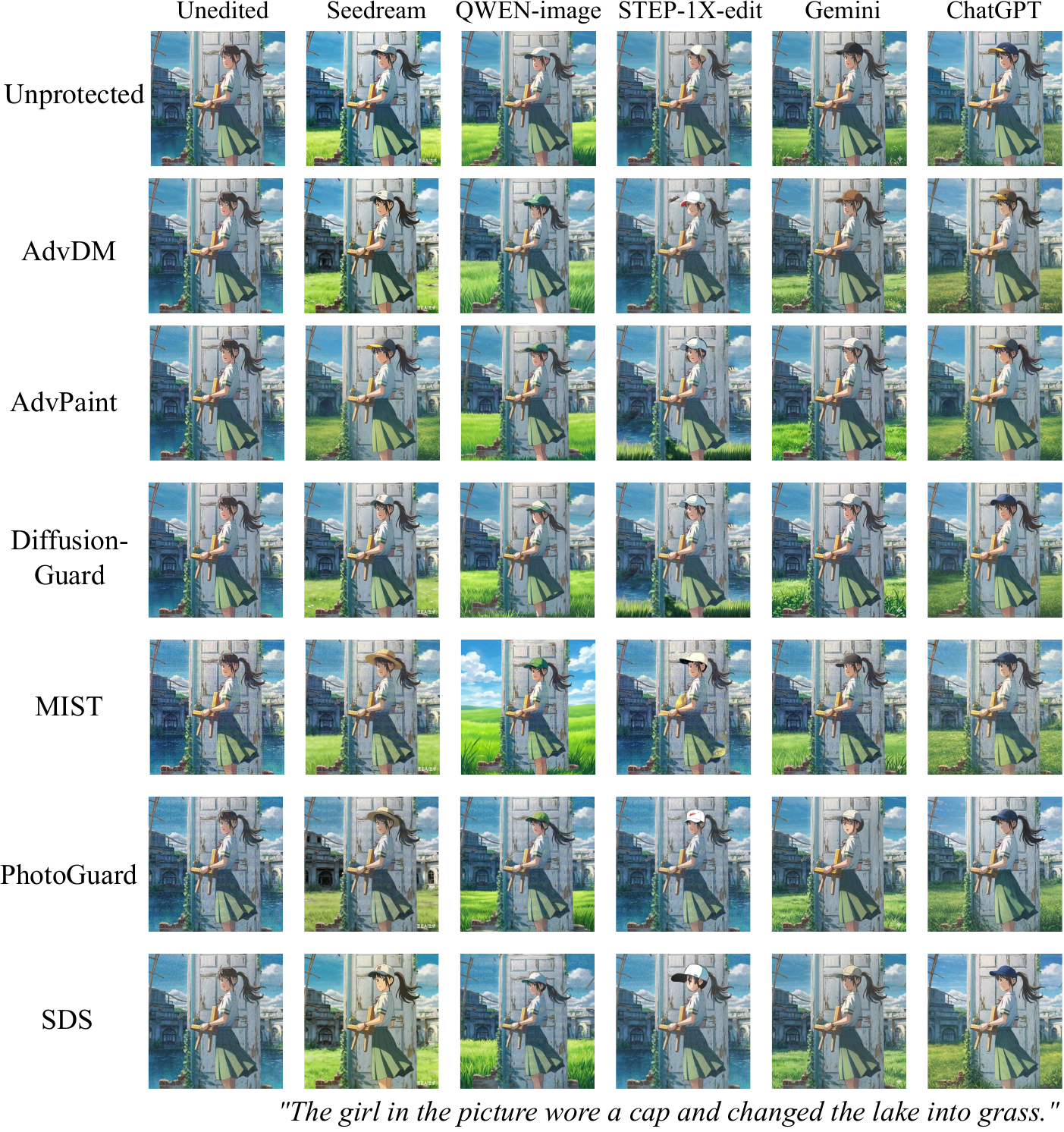}
\caption{Extended qualitative examples complementing Figure~\ref{fig:cross_arch_llm} with a more complex cartoon image and instruction. Rows show clean and protected inputs (six defenses); columns show edited results from Seedream (Doubao), Qwen-Image, Step1X-Edit, Gemini Pro, and ChatGPT-4o.}
\label{fig:cross_arch_llm_appendix}
\end{figure*}

\begin{figure*}[t]
\centering
\includegraphics[width=\textwidth]{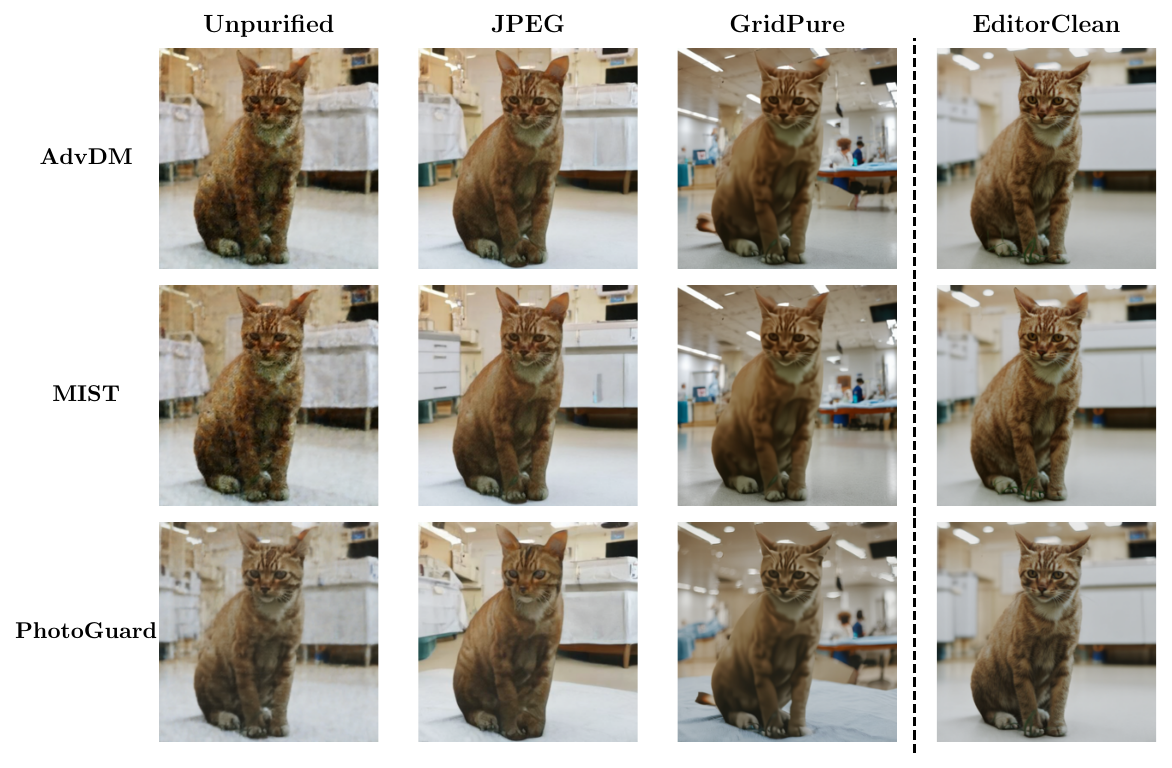}
\caption{Qualitative downstream editing results on Step1X-Edit (DiT), corresponding to Table~\ref{tab:step1x_results}. }
\label{fig:step1x_results_visual}
\end{figure*}

\begin{figure*}[t]
\centering
\includegraphics[width=\textwidth]{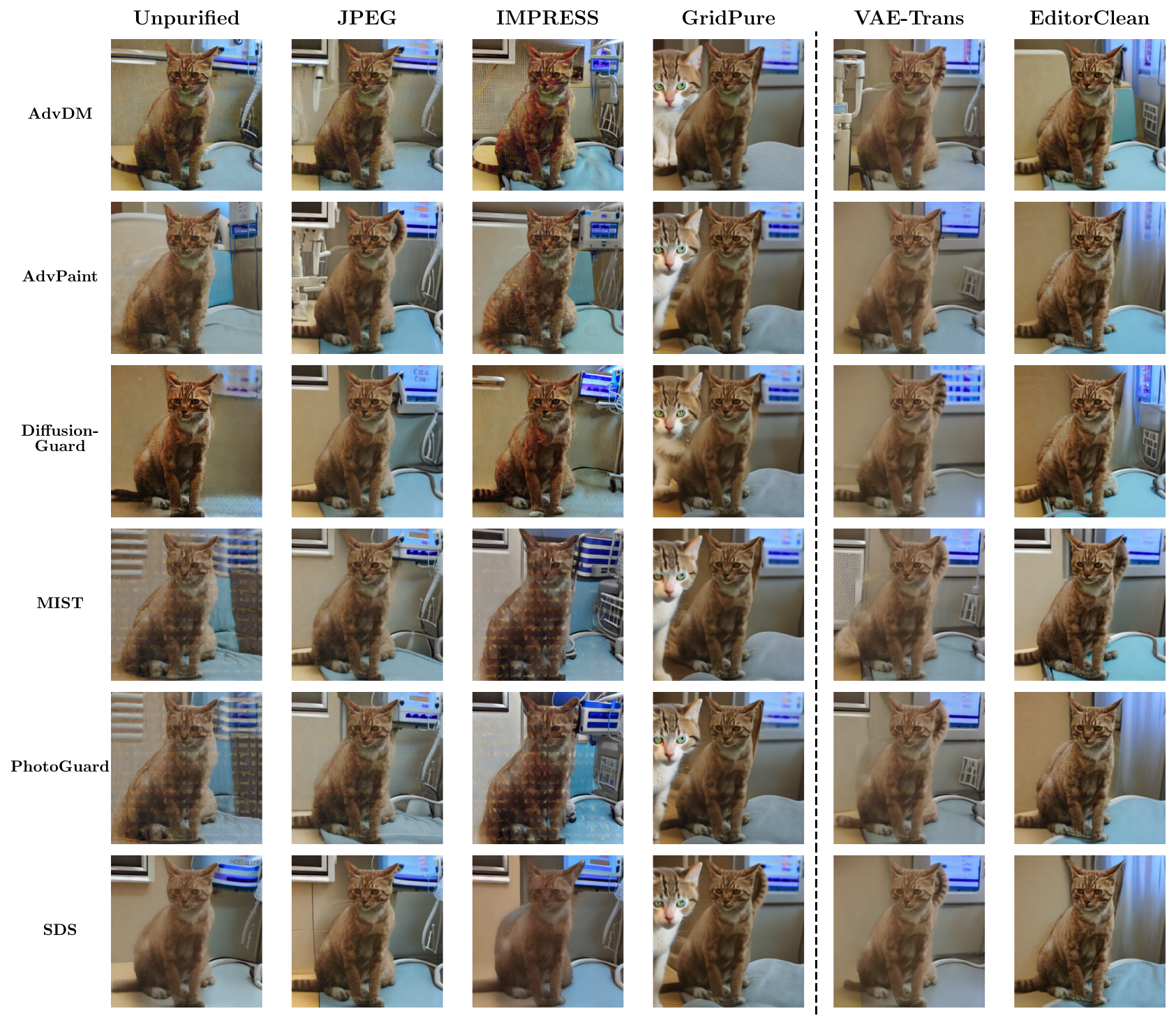}
\caption{Qualitative visualization of the perturbation budget ablation ($\epsilon=8/255$) in Table~\ref{tab:epsilon_ablation}.}
\label{fig:epsilon_ablation_visual}
\end{figure*}

\begin{figure*}[t]
\centering
\includegraphics[width=0.9\textwidth]{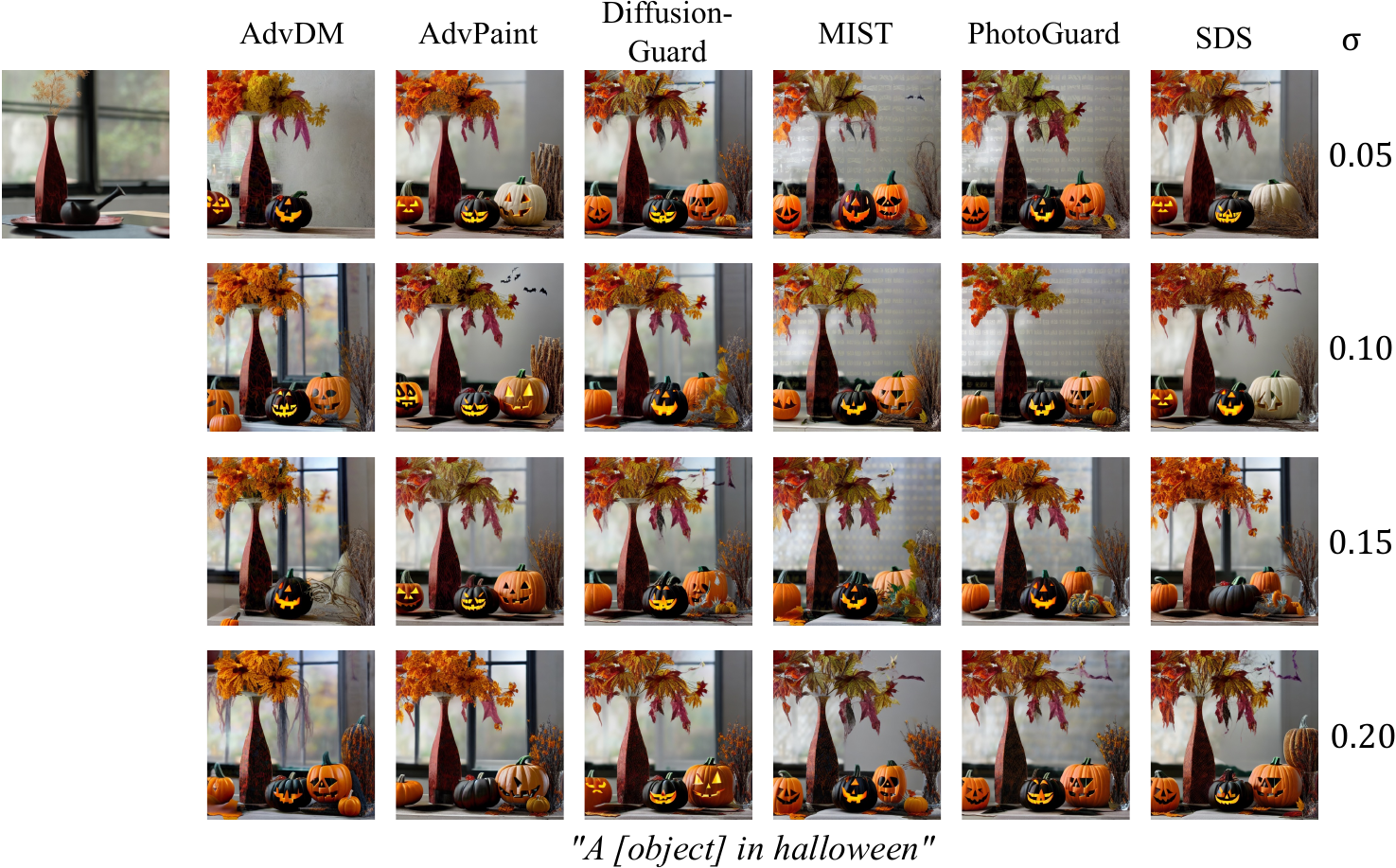}
\caption{Qualitative visualization of the Gaussian noise strength ablation in Table~\ref{tab:gaussian_ablation}.}
\label{fig:gaussian_ablation_visual}
\end{figure*}

\begin{figure*}[t]
\centering
\includegraphics[width=0.9\textwidth]{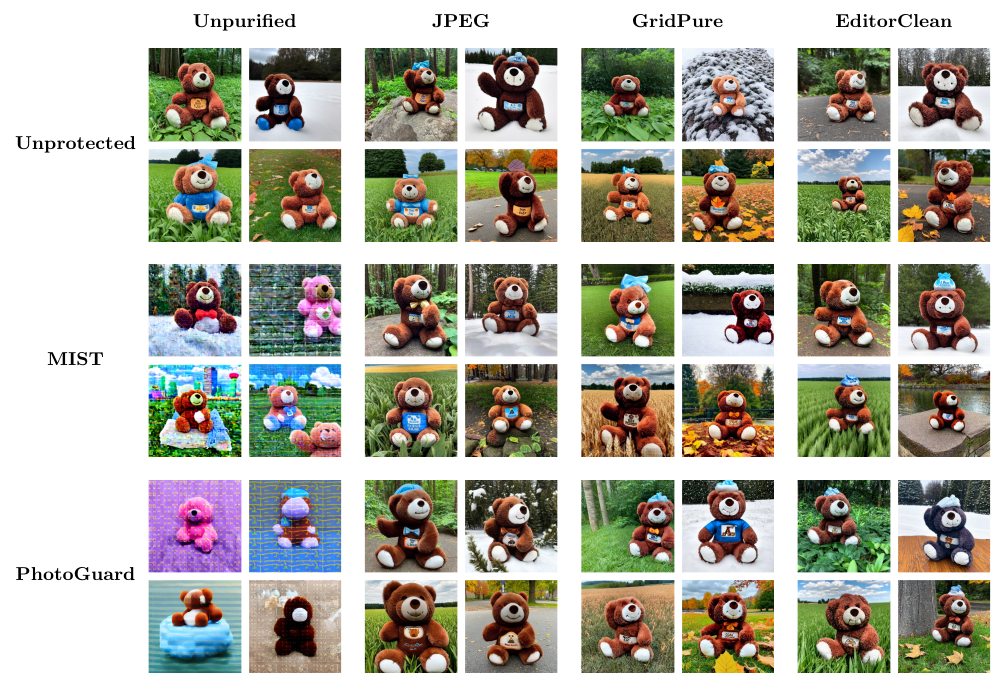}
\caption{Qualitative DreamBooth fine-tuning results for \texttt{bear plushie}, corresponding to Table~\ref{tab:dreambooth_results}.}
\label{fig:dreambooth_results_visual}
\end{figure*}

\begin{figure*}[t]
\centering
\includegraphics[width=\textwidth]{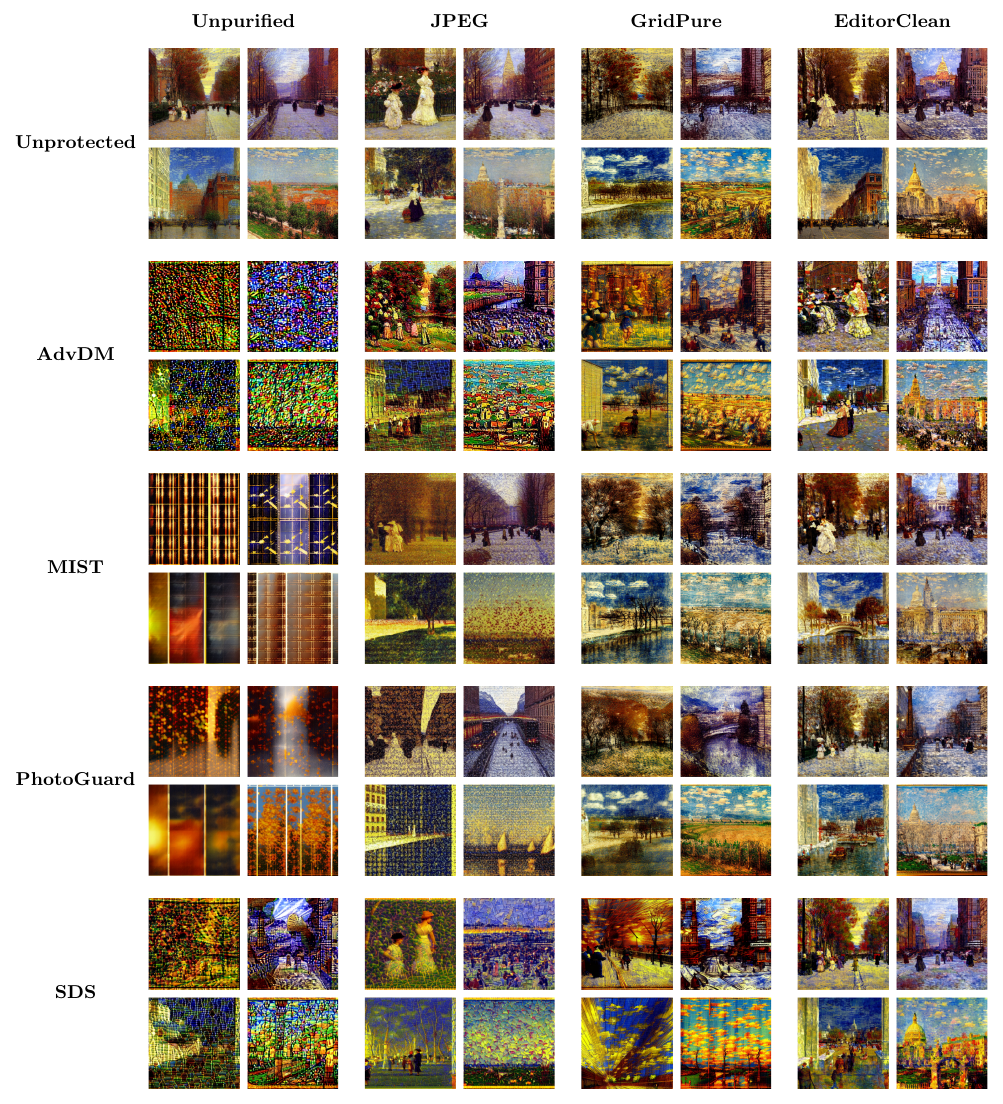}
\caption{Qualitative textual inversion style mimicry results for \textit{Childe Hassam}, corresponding to Table~\ref{tab:textual_inversion_results}.}
\label{fig:textual_inversion_results_visual}
\end{figure*}

\section{Dataset Details}
\label{sec:dataset_details}

This section provides a detailed description of the DiffusionGuard dataset~\cite{choi2024diffusionguard} used for all evaluations in Section~\ref{sec:experiments}.

\noindent\textbf{Dataset Composition.}
The DiffusionGuard dataset is specifically designed to assess the robustness of image protection methods against unauthorized editing.
It comprises 42 test images spanning three semantic categories: (i)~32 celebrity portraits, (ii)~5 inanimate objects, and (iii)~5 animal images.
The celebrity portraits include diverse subjects with varying lighting conditions, backgrounds, and facial attributes to ensure comprehensive evaluation.
The object and animal categories provide complementary test cases beyond human subjects, enabling assessment of protection generalization across semantic domains.

\noindent\textbf{Task Construction.}
For each of the 42 images, the dataset provides 10 text-guided editing instructions and 5 diverse inpainting masks.
The editing instructions cover four major categories: (i)~attribute modification, (ii)~object replacement, (iii)~style transfer, and (iv)~background alteration.
The 5 inpainting masks per image are carefully designed to target different spatial regions and mask sizes, ranging from localized edits to large-area modifications.
By pairing each of the 42 images with 10 instructions and 5 masks, the dataset yields a total of $42 \times 10 \times 5 = 2{,}100$ unique editing tasks, providing comprehensive coverage for statistically meaningful evaluation.

\section{Hyperparameters for Unlearnable Sample Experiments}
\label{sec:unlearnable_hyperparams}

This section provides the hyperparameter settings for the unlearnable sample experiments in Section~\ref{sec:unlearnable}.

\noindent\textbf{DreamBooth Personalization.}
We fine-tune Stable Diffusion v1.5 using the official Diffusers training script with learning rate $10^{-6}$ for 500 steps.
After fine-tuning, we generate 380 images using diverse prompts.
All protected images use a reduced perturbation budget of $\epsilon = 8/255$.

\noindent\textbf{Textual Inversion Style Mimicry.}
For each of the three artists, we use seven training images and learn a style embedding token.
Training uses a learning rate of $1 \times 10^{-4}$ for 3,000 steps.
We then generate 100 images per artist using the prompt template ``a painting in the style of \texttt{<style>}'', where \texttt{<style>} denotes the learned embedding.
All protected images use a reduced perturbation budget of $\epsilon = 8/255$.

\end{document}